\documentclass[pra,twocolumn,aps,superscriptaddress,longbibliography]{revtex4-1}

\usepackage[normalem]{ulem}
\usepackage{amsmath}
\usepackage{algorithm}
\usepackage{algorithmic}
\usepackage{amssymb}
\usepackage{lineno}
\usepackage{graphicx}
\usepackage[usenames,dvipsnames]{color}
\usepackage{braket}
\usepackage{bm}
\usepackage{mathtools}
\usepackage{times}
\usepackage{hyperref}
\usepackage{pdfpages}
\usepackage{float}
\usepackage{lipsum}
\usepackage[dvipsnames]{xcolor}
\hypersetup{
  colorlinks=true,
  citecolor=blue,
  linkcolor=blue,
  urlcolor=blue}
\usepackage{tikz}
\usetikzlibrary{decorations.markings}
\usetikzlibrary{decorations.pathmorphing,snakes}
\usetikzlibrary{arrows.meta}
\tikzset{middlearrow/.style={
        decoration={markings,
            mark= at position 0.5 with {\arrow{#1}} ,
        },
        postaction={decorate}
    }
}

\makeatletter
 \AtBeginDocument{\let\LS@rot\@undefined}
\makeatother

\begin{document}
\title{Exact diagonalization using hierarchical wave functions and 
       calculation of topological entanglement entropy}

\author{Deepak Gaur}
\affiliation{Physical Research Laboratory,
             Ahmedabad - 380009, Gujarat,
             India}
\affiliation{Indian Institute of Technology Gandhinagar,
             Palaj, Gandhinagar - 382355, Gujarat,
             India}

\author{Hrushikesh Sable}
\altaffiliation{Present address: Department of Physics, Virginia Tech,
             Blacksburg, Virginia 24061, USA}
\affiliation{Physical Research Laboratory,
             Ahmedabad - 380009, Gujarat,
             India}

\author{D. Angom}
\affiliation{Department of Physics, Manipur University,
             Canchipur - 795003, Manipur,
             India}

\begin{abstract}
In this work we describe a new technique for numerical exact diagonalization.
The method is particularly suitable for cold bosonic atoms in optical lattices,
in which multiple atoms can occupy a lattice site. We describe the use of the 
method for Bose-Hubbard model of a two-dimensional square lattice system as 
an example; however, the method is general and can be applied to other lattice
models and can be adapted to three-dimensional systems. The proposed numerical 
technique focuses in detail on how to construct the basis states as a hierarchy
of wave functions. Starting from single-site Fock states, we construct the 
basis set in terms of row states and multirow states. This simplifies the 
application of constraints and calculation of the Hamiltonian matrix. 
The approach simplifies the calculation of the reduced density matrices, and 
this has applications in characterizing the topological entanglement of the
state. Each step of the method can be parallelized to accelerate the 
computation. As a case study, we discuss the computation of the spatial 
bipartite entanglement entropy in the correlated $\nu =1/2$ fractional quantum
Hall state.
\end{abstract}

\maketitle


\section{Introduction}

Very few problems in quantum many-body systems are analytically solvable. 
One often resorts to numerical techniques to gain insights on these systems.
Numerical mean-field methods are straightforward and easy to use
\cite{chaikin_95}; however, they fail to accommodate the correlations and
entanglement of a quantum system. The quantum Monte Carlo (QMC)-based methods
like the stochastic series expansion \cite{sandvik_91, sandvik_92} are powerful
numerical techniques. These are often used to study the properties of the
system in thermodynamic limit. However, for fermionic systems, QMC suffers from
the ``sign problems" \cite{henelius_00}. In addition, a direct access to the 
full quantum state is not possible with QMC. To obtain the complete information
of the system, the numerical exact-diagonalization (ED) is accurate and 
dependable. As the name suggests, the Hamiltonian matrix is ``exactly" 
diagonalized after choosing an appropriate basis, and generating the 
Hamiltonian matrix with the basis states.For most of the studies on 
quantum-mechanical systems at low temperatures, the low-energy states, namely, 
the ground state and few excited states, are sufficient to describe the 
important properties. This simplicity allows the usage of the Lanczos algorithm
for faster numerical diagonalization \cite{lanczos_1952, calvetti_94}. 
Although ED offers the advantage of access to the exact solution, it is 
however, computation intensive and grows with complexity of the problem. 
For instance, for spins on a lattice, the Hilbert space of the system grows 
exponentially with the lattice size. Thus, ED limits the system size that one 
can study. In the literature there exists large-scale ED studies for spin-$1/2$
kagome Heisenberg antiferromagnet with $\approx 40-50$ spins
\cite{lauchli_11, lauchli_19}, bosonic fractional quantum Hall on a 
$12\times 4$ lattice \cite{bai_18}.

 In this work, we introduce an ED-based method well suited for ultracold bosons
in optical lattices. These quantum systems serve as excellent quantum
simulators to study various condensed-matter systems \cite{lewenstein_12}, as
they allow remarkable control and tunability of system parameters in
experiments \cite{greiner_02,lewenstein_07}. The Bose-Hubbard model (BHM) is a
prototypical model which describes the low-energy physics of these systems
\cite{hubbard_63,fisher_89,jaksch_98}. The mean-field methods like the
Gutzwiller mean-field theory \cite{rokhsar_91, sheshadri_93, oktel_07} provide
qualitative insights, but do not represent the correlations accurately. 
One area where ED has provided detailed insights and reliable results is in 
identifying  topological phases like the quantum Hall states. Such states
can be experimentally realized with the ultracold atoms by introducing 
synthetic gauge fields \cite{aidelsburger_11,aidelsburger_13,garcia_12}. 
Mean-field methods are not suitable for studying these strongly correlated 
phases. There are various proposals to realize bosonic analogues of fractional
quantum Hall (FQH) states \cite{dalibard_2011, aidelsburger_2018, 
cooper_2019, hauke_2022}. Recently, a $\nu = 1/2$ bosonic FQH state was 
experimentally observed with ultracold atoms \cite{leonard_23}. A rudimentary 
version of the proposed ED method, as a component of the cluster Gutzwiller 
mean-field theory implementation, was employed in our previous works to 
analyze the quantum phases in an optical lattice for a wide variety of systems.
These include bosonic quantum Hall phase \cite{bai_18,bai_19,gaur_22}, 
Bose-glass phase \cite{pal_19}, supersolid phase in dipolar bosons 
\cite{suthar_20, suthar_20_2}, dipolar bosonic mixtures \cite{bai_20}, 
dipolar bosons in multilayer optical lattice \cite{bandyopadhyay_22}.

 The ED method we have developed relies on a hierarchical sequence of basis
states. The single-site Fock states constitute the lowest hierarchy of the
sequence. Next, the tensor products of these states define a 1D or row state.
Finally, the tensor product of the row states, referred to as the multirow 
states, form the basis set of the ED method. This approach facilitates
applying constraints on the system, such as restrictions on the occupancies in 
the rows and system as a whole \cite{bai_18}. Once the basis states, also
referred interchangeably as cluster states, are generated, calculation of the
Hamiltonian matrix is the next step in ED. For lattice systems the Hamiltonian
matrix is, in general, sparse. We take advantage of this feature and calculate
only the nonzero elements by identifying the nonzero elements corresponding
to a cluster state. This, however, requires a more sophisticated approach when
constraints are included. We then develop and present appropriate schemes to
calculate bipartite entanglement entropy, which requires partitioning the
system into two distinct subsystems. It is a good measure of correlations in a
quantum state \cite{kitaev_06, levin_06} and valuable in studying quantum Hall
states.

 We have organized the remainder of this article as follows.
In Sec. {\ref{sec_ed_full}} we describe our implementation of the ED technique
in terms of row states as the fundamental building blocks. We describe the
construction of the set of basis states and that of the Hamiltonian in this
section. This is followed by Sec. {\ref{sec_statred}}, where we discuss the
modified construction procedures when additional constraints are imposed for
filtering out the basis states. These constraints reduce the size of the
Hilbert space. In Sec. {\ref{sec_bhm_res}} we establish the validity of our 
ED method and benchmark it by calculating selected ground-state properties of
BHM. In Sec. {\ref{sec_entang}} we demonstrate the usage of the algorithm for 
studying the spatial bipartite entanglement in the $\nu=1/2$ bosonic FQH state.
Finally, we conclude in Sec. {\ref{sec_discussions}}.


\section{Complete basis set } \label{sec_ed_full}

Consider a system of ultracold bosonic atoms loaded in a two-dimensional (2D) 
optical lattice and to simplify the description, consider the geometry to be 
square. Such a system is well described by the Bose-Hubbard Hamiltonian given by
\begin{eqnarray}
   \hat{H}_{\text{BHM}} = \sum_{p, q}\bigg [ &&-J\Big(
             \hat{b}^{\dagger}_{p+1, q}\hat{b}_{p, q} 
             + \hat{b}^{\dagger}_{p, q+1}\hat{b}_{p, q} + {\rm H.c.}\Big)
             \nonumber\\ 
             &&+ \frac{U}{2} \hat{n}_{p, q} (\hat{n}_{p, q}-1) \bigg], 
\label{ham}  
\end{eqnarray}
where $p$ $(q)$ is the lattice site index along the $x$ $(y)$ direction, 
$\hat{b}_{p,q}$ ($\hat{b}_{p,q}^{\dagger}$) is the annihilation (creation)
operator at lattice site $(p,q)$, $\hat{n}_{p,q}$ is the number operator, and
$U$ is the on-site interaction strength. The eigenstate of a system with a
size of $K\times L$ can be written as the linear combination,
\begin{equation}
   \ket{\Psi} = \sum_{\substack{n_{1,1} \\ \cdots\\ n_{p,q}\\ \cdots\\ 
                n_{K,L}}} C_{ n_{1,1},\cdots n_{p,q},\cdots n_{K,L}}
                \ket{ n_{1,1},\cdots n_{p,q},\cdots n_{K,L}},
\label{wavefunc}  
\end{equation}
where $n_{p,q}$ is the occupancy at the lattice site $(p,q)$, and $K$ ($L$)
is the number of lattice sites along the $x$ ($y$) direction. To obtain 
the ground state or any other excited states, the coefficients 
$C_{n_{1,1},\cdots n_{p,q},\cdots n_{K,L}}$ corresponding to the basis state 
$\ket{ n_{1,1},\cdots n_{p,q},\cdots n_{K,L}}$ must be defined. 
The basis states are defined in terms of a hierarchy of multiple lattice
sites, with the starting point being the single-site Fock states.


\subsection{Construction of the basis states}

As mentioned earlier, the occupancy of a lattice site $(p,q)$ is denoted by 
$n_{p,q}$, and the corresponding Fock state can be represented as 
$\ket{n_{p,q}}$. Although $n_{p,q}$ can in principle be any non-negative 
integer, let us consider only $N_B$ possible choices 
$n_{p,q} \in \{0, 1, 2,\cdots, N_B -1\}$. A configuration of occupancies of 
lattice sites along the $q$th row can then be collectively represented by a 
``row state", defined as $\ket{n_{1,q},n_{2,q},\cdots n_{p,q},\cdots n_{K,q}} 
\equiv \prod_{p = 1}^{K} \otimes \ket{n_{p,q}}$, where
$1\leqslant p\leqslant K$ is the lattice site index along the row. In general, 
the label $q$ corresponding to the row can be omitted, as it is common for all 
the lattice sites along the row. And a row state can be defined as
\begin{equation}
   \ket{ n_{1},n_{2},\cdots n_{p},\cdots n_{K}} 
   = \prod_{p = 1}^{K} \otimes \ket{n_p}.
\label{rowstates}  
\end{equation}
The advantage of defining the row states is that any configuration of particles
in the entire lattice can be expressed as a direct product of the row states. 
Thus the row states are the building blocks of the basis states of the entire 
lattice system.

 The occupancies, as mentioned earlier, can be 
$0 \leqslant n_p \leqslant N_B - 1$. Then the total number of possible 
row-state configurations is $\alpha = (N_B)^K$. Each row-state configuration 
is uniquely identified by the quantum number $m$ and represented as
\begin{equation}
   \ket{\phi_m} \equiv \prod_{p = 1}^{K} \ket{n_p},
\label{phi}
\end{equation}
with $m \in \{0,1,2,\cdots \alpha-1\}$. The quantum number $m$ and the number
of particles ($N_m$) in the row state are given by
\begin{equation}
   m = \sum_{p=1}^{K} n_p \,N_B^{p-1}, \text{and }
   N_m = \sum_{p=1}^{K} n_{p}.
\label{n-row}
\end{equation}
Using the row states, we construct multirow states, which are the occupancy
configurations in multiple rows. The multirow states of two rows, referred
as two-row states, is represented as $\ket{\Phi^2}$. In the present case, the
lattice dimension of $\ket{\Phi^2}$ is $(K \times 2)$, 
and each of the two-row states is a direct product of the row states 
$\ket{\phi_m}$, defined in Eq.~(\ref{phi}) as
\begin{equation}
   \Ket{\Phi^2_{M}} 
                  = \ket{\phi_{m_2}} \otimes \ket{\phi_{m_1}}.
\end{equation}
Here, the label $M$ corresponds to a two-dimensional vector 
${\bf M} = (m_1, m_2)$. It contains information about the row states which
constitutes the two-row state. Thus, the total number of possible two-row 
states is  $\alpha^2$, and for a combination of $m_1$ and $m_2$, the 
corresponding state label is $M=\alpha(m_1) + m_2$. Similarly, multirow states 
of a higher number of rows can be constructed from the direct product of 
multirow states with fewer number of rows and row states, as 
\begin{eqnarray}
   \Ket{\Phi^3_{M'}} 
                &&= \Ket{\Phi^2_{M}} \otimes \ket{\phi_m} \,,
                         \nonumber \\
   \Ket{\Phi^4_{M'}}
                &&= \Ket{\Phi^3_{M}} \otimes \ket{\phi_m} \,.
\label{cluster_state}
\end{eqnarray}
Considering the hierarchy of the multirow states, different rank multirow 
states with $q$ rows $\Ket{\Phi^q_{M}}$ is identified by a vector 
${\mathbf M} = (m_1, m_2, \cdots m_q)$ in a $q$-dimensional space. The 
row-state quantum numbers are the components of the vector, and each axes 
represents one of the row states. Various $q$-row states are assigned 
a unique integer index $M$ corresponding to the vector label ${\mathbf M}$,
and the number of particles in the state $N_M$ are
\begin{equation}
	M = \sum_{j=1}^{q} m_j (N_B)^{K(L-j)}, \text {and }
   N_M = \sum_{p=1}^{K}\sum_{q'=1}^{q}n_{p,q'}.
 \label{int_label}
\end{equation}

\begin{figure}[t]
   \begin{center}
        \includegraphics[width=8.5cm]{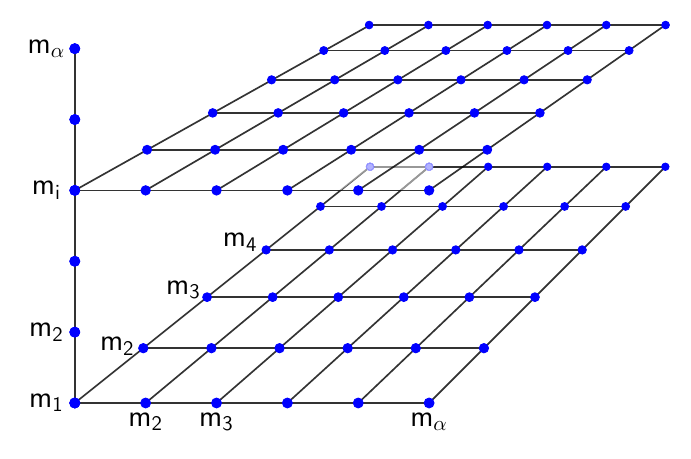}
	   \caption{Schematic illustration of the hyper-space
                spanned by the row-state quantum numbers along various rows 
                which together constitute the lattice. The basis state are 
                identified by the grid points in the three-dimensional space 
                for a lattice consisting of three rows.}
   \label{basis_lattice}
   \end{center}
\end{figure}

Generalizing the definition of multirow states, the basis states of a system 
with $L$ rows and $K$ columns is constructed as a direct product of two
appropriately chosen multirow states. Each of the basis states 
$\Ket{\Phi^L_{M}}$ is identified by an $L$-dimensional vector 
${\mathbf M} = (m_1, m_2, \cdots m_L)$ 
and an index $M$. For example, the basis for a $K \times 7$ lattice system 
can be constructed as a direct product of multirow states 
$\Ket{\Phi^4_{M_1}}$ and $\Ket{\Phi^3_{M_2}}$ as given below:
\begin{equation}
   \Ket{\Phi^7_{M}}
                = \Ket{\Phi^4_{M_1}} \otimes \Ket{\Phi^3_{M_2}}.
\label{basis}
\end{equation}
The basis states are uniquely identified by a vector ${\bf M}$ and are
labeled with an integer $M$ according to relation given in 
Eq.~(\ref{int_label}). Thus, in general the basis states of $K\times L$ can
be defined as the direct product
\begin{equation}
   \Ket{\Phi^L_{M}}
                = \Ket{\Phi^{L_1}_{M_1}} \otimes \Ket{\Phi^{L_2}_{M_2}},
  \label{phi_genrl}
\end{equation}
where $L=L_1 + L_2$. Below is a schematic representation showing the 
occupancies over the lattice which constitutes a basis state, being identified 
as a column vector of row states, which is the essence of Eq.~(\ref{basis}):
\begin{equation}
  \begin{pmatrix}
     n_{1,L}  & n_{2,L}  & \cdots & n_{p,L}  & \cdots & n_{K, L} \\
     \vdots   & \vdots   &        & \vdots   &        & \vdots   \\
     n_{1,q}  & n_{2,q}  & \cdots & n_{p,q}  & \cdots & n_{K,q}  \\
     \vdots   & \vdots   &        & \vdots   &        & \vdots   \\
     n_{1,2}  & n_{2,2}  & \cdots & n_{p,2}  & \cdots & n_{K,2}  \\
     n_{1,1}  & n_{2,1}  & \cdots & n_{p,1}  & \cdots & n_{K,1}
  \end{pmatrix}
  \equiv
  \begin{pmatrix}
     & m_L   \\
     & \vdots  \\
     & m_q  \\
     & \vdots  \\
     & m_2  \\
     & m_1 
  \end{pmatrix}.
  \nonumber
\label{matrix}
\end{equation}
Thus any basis state representing the configuration of particles on a 2D
lattice with $L$ rows can be thought of as a point in the $L$-dimensional 
hyper-space spanned by the row states $\ket{\phi_m}$, as shown in 
Fig.~\ref{basis_lattice}.
With these definitions, the number of possible basis states of the system 
is $\alpha^L$. Using these, any state of the system can be expressed in
terms of the basis states as
\begin{equation}
   \ket{\psi} = \sum_{M = 0}^{\alpha^L -1} 
		C^ M \Ket{\Phi^L_{M}},
\end{equation}
where $C^M$ is the coefficient corresponding to the basis $\Ket{\Phi^L_{M}}$ 
in the wave function $\ket{\psi}$. The corresponding vector 
${\bf M} = (m_1,m_2,\cdots,m_L)$ has the information regarding the occupancies 
along all the rows corresponding to $M$th basis state.

\subsection{Construction of Hamiltonian matrix}
\label{sect-ham-mat}

Let us now consider the BHM Hamiltonian given in Eq.~(\ref{ham}). Consider the
hopping term $-J \hat{b}^{\dagger}_{p+1, q}\hat{b}_{p, q}$, which describes the
hopping of a particle from the lattice site $(p, q)$ to $(p+1, q)$. The 
action of this term on a basis state affects only two lattice sites of the
$q$th row. Thus, its operation on the row yields
\begin{eqnarray}
   &&\hat{b}^{\dagger}_{p+1, q} \hat{b}_{p, q}
     \ket{n_1, \cdots n_p, n_{p+1}, \cdots n_{K}} 
     \nonumber \\
   &&= \sqrt{n_p\,(n_{p+1} +1)}
         \ket{n_1, \cdots (n_p -1), (n_{p+1} +1),\cdots n_{K}} \,,
         \nonumber 
\end{eqnarray}
if the hopping is allowed, which corresponds to $n_p -1 \geqslant 0$ and
$n_{p+1} \leqslant N_B -1$. Otherwise, the action of the hopping term results
to zero. Thus for a possible hopping, this hopping term only modifies the
row state $\ket{\phi_{m_q}}$ to a different row state $\ket{\phi_{m'_q}}$
\begin{equation}
   \hat{b}^{\dagger}_{p+1, q}\hat{b}_{p, q} \ket{\phi_{m_q}} 
       = \sqrt{n_p\,(n_{p+1} +1)} \ket{\phi_{m'_q}},
\label{hop_phi}
\end{equation}
and the quantum numbers of the row states are related as
\begin{eqnarray}
   m'_q - m_q &=& \sum_{j=1}^{K} (n'_{j}- n_{j}) \,N_B^{j-1},
                \nonumber \\
              &=& N_B^{p-1} \big( N_B -1 \big).
\label{diff_phi}
\end{eqnarray}
Thus, for a possible $x$ hopping, the operation of the hopping term
$-J \hat{b}^{\dagger}_{p+1, q}\hat{b}_{p, q}$ on the basis state 
$\Ket{\Phi^L_{M}}$ can be written as
\begin{equation}
   -J \hat{b}^{\dagger}_{p+1, q}\hat{b}_{p, q} \Ket{\Phi^L_{M}}
               = -J \sqrt{n_{p,q}\,(n_{p+1,q} +1)} \Ket{\Phi^L_{M'}},
\label{hop-Phi}
\end{equation}
where the corresponding basis states vectors ${\bf M} = (m_1, m_2, \cdots m_L)$
and ${\bf M'} = (m'_1, m'_2, \cdots m'_L)$ are related as
\begin{equation}
   m'_j = \begin{cases}
           m_j  & \text{if }  j \ne q,  \\
           m_q + N_B^{p-1} \left ( N_B -1 \right ) & \text{if } j=q.
          \end{cases}
\label{diff-Phi}
\end{equation}
Here, $n_{p,q}$ is the occupancy at site $(p,q)$ for the basis 
$\ket{\Phi^L}_{\bf M}$. Thus we can calculate the nonzero matrix element of
the hopping term as 
\begin{eqnarray}
  (-J \hat{b}^{\dagger}_{p+1, q}\hat{b}_{p, q}) _{M' M}
		    &&\equiv \Bra{\Phi^L_{M'}}
                    -J \hat{b}^{\dagger}_{p+1, q}\hat{b}_{p, q}
		    \Ket{\Phi^L_{M}},  \nonumber \\
                    &&= -J \sqrt{n_{p,q}\,(n_{p+1,q} +1)} .
\end{eqnarray}
Similarly, consider the hopping term
$-J \hat{b}^{\dagger}_{p, q+1}\hat{b}_{p, q}$, which describes the hopping of
particles from lattice site $(p,q)$ to $(p,q+1)$ between two neighboring rows.
The hopping term will modify the row states at $q+1$ and $q$th rows.
Thus, for a possible $y$ hopping, the action of the hopping term
$-J \hat{b}^{\dagger}_{p, q+1}\hat{b}_{p, q}$ on the basis state
$\Ket{\Phi^L_{M}}$ is given by
\begin{equation}
 -J \hat{b}^{\dagger}_{p, q+1}\hat{b}_{p, q} \Ket{\Phi^L_{M}}
                      = -J \sqrt{n_{p,q}\,(n_{p,q+1} +1)} \;
                        \Ket{\Phi^L_{M'}},
\end{equation}
where
\begin{equation}
   m'_j = \begin{cases}
           m_j                 & \text{if } j \ne q, j \ne q+1,  \\
           m_q - N_B^{p-1}     & \text{if } j = q,               \\
           m_{q+1} + N_B^{p-1} & \text{if } j = q+1,
          \end{cases}
\label{yhop_mat}
\end{equation}
and this contributes to the $( M',  M)$ Hamiltonian matrix element. In a 
similar way, the other terms of the Hamiltonian can be evaluated.


\section{State Reduction} \label{sec_statred}

The row states $\ket{\phi_m}$ and multirow states $\Ket{\Phi^q_{M}}$ 
considered so far are without restrictions on the total number of particles
$N=\sum_{p=1}^{K}\sum_{q=1}^{L} n_{p,q}$, except for the constraint of 
$n_{p,q}$ not exceeding $N_B-1$. Thus, the basis states described by 
Eq.~(\ref{phi_genrl}) include all possible states with 
$0\leqslant N \leqslant (N_B-1)KL$. Such a basis set is not appropriate 
for calculations with fixed $N$ or in studies using the canonical ensemble 
for BHM-like Hamiltonians, where the number operator commutes with the
Hamiltonian and is thus a conserved quantity. For such cases, let us refine the
basis-state construction to impose the constraint of selecting states with a
particular value of $N$. This implies  that $n_{p,q}$ can range from $0$ to
$\min(N, N_B-1)$. Depending on the problem of interest, additional constraints
can be imposed on basis states. For example, at unit filling $(N= K\times L)$
average occupancies at each lattice site is $1$, and thus basis states with
occupancies $n_{p,q}>2$ may not contribute to the lowest energy sector. So it
is appropriate to impose constraints, which reduces the number of the basis
states. These considerations, however, require modifications in the basis
construction procedure and Hamiltonian matrix calculation.


\subsection{Construction of the row states}

The single-site Fock states $|n_{p,q}\rangle$ considered so far have
occupancies $0\leqslant n_{p,q}\leqslant (N_B-1)$. For quantum states with 
higher average occupancies, it is appropriate to modify the choice of 
$|n_{p,q}\rangle$ with the constraint 
\begin{equation}
  \eta \leqslant n_{p,q}\leqslant (N_B-1).
\label{ss_bound}
\end{equation}
This fixes the number of single-site Fock states to $N_B -\eta$ but modifies 
the lowest possible occupancy of the single-site Fock state to $\eta$. 
The other constraint is to restrict the number of particles in a row state
within a range of values,
\begin{equation}
   \sigma \leqslant \sum_{p=1}^{K} n_{p,q} \leqslant \sigma +\delta , 
\label{row_bound}
\end{equation}
where $\sigma \geqslant \eta K$ and $\sigma + \delta \leqslant (N_B-1)K$.
This sets a limit on the minimum number of particles in a row state as
$\sigma$ and fixes the range to $\delta$. Such a constraint is appropriate
to generate basis sets of optimal sizes to study quantum phases with low 
average occupancies. Further, as the constraint is on the range of $N_m$,
it can also be imposed to limit the basis set size but ensure that it 
represents the number fluctuation of the quantum phase appropriately. 
Thus, Eq.~(\ref{phi}) can still be used to describe the row states with the 
constraints in Eqs.~(\ref{ss_bound}) and (\ref{row_bound}). 
Due to the state-reduction constraints set by
Eq.~(\ref{row_bound}), the quantum number $m$ of the selected row states 
need not be consecutive integers. And thus $m$ is unsuitable as index 
to identify row states and labeling of the Hamiltonian matrix elements. 
Consider the set of the selected row-state quantum numbers in ascending order, 
$\mathcal{S} = \{m_i: m_i < m_{i+1} \;\text{and}\; i, i+1 \in\mathcal{S}'\}$.
Here, $i$ is the sequence label, which forms an index set
$\mathcal{S}' = \{ 1,2, \cdots \beta \}$, and $\beta$ is the number of the
selected row states. As mentioned earlier, $m_i$ and $m_{i+1}$ need not be
consecutive integers but are consecutive elements of $\mathcal{S}$.
For better organization and tractability, we redefine the row states with the
label $i$ denoted as $\ket{\phi}_{i}$. This redefinition implies 
$\ket{\phi}_{i} \equiv \ket{\phi_{m_i}}$. Appendix \ref{appendA} illustrates, 
as an example, the generation of the row-state configurations for
appropriately chosen state reduction constraints. As described earlier, the 
row states are used to construct the basis for the entire system. To generate 
the basis set, only the multirow states satisfying the constraints are 
selected. The selection is simplified by the use of the multirow states, which
forms the building block for the construction of the basis states.


\subsection{Multirow states with constraints}

  The constraints as expressed in Eqs.~(\ref{ss_bound}) and 
(\ref{row_bound}) are for the single site and row state, respectively. For
the entire lattice system, a constraint on $N$ can be imposed as
\begin{equation}
   \Sigma \leqslant \sum_{p=1}^{K} \sum_{q=1}^{L} n_{p,q} 
   \leqslant \Sigma +\Delta, 
  \label{cluster_bound}
\end{equation}
where, $\Sigma \geqslant \sigma L$ and $\Delta \leqslant \delta L$.
Like in the case of a row state, this sets a limit on the minimum number of 
particles in a basis state as $\Sigma$ and fixes the range to $\Delta$.
This constraint is suitable when total number of particles in the lattice is 
not fixed. Thus, to reduce the basis set to an optimal size but appropriate to
describe the quantum phase of interest, constraints in Eqs.~(\ref{ss_bound}),
(\ref{row_bound}), and (\ref{cluster_bound}) are imposed to the single site,
row state, and basis state, respectively. It must, however, be mentioned that
for studies with the micro-canonical ensemble, the total number of particles 
is fixed and corresponds to $\Delta = 0$.

As described earlier, the  multirow states consisting of $q$ rows
are denoted by $\Ket{\Phi^q_{M}}$, with a corresponding $q$-dimensional
vector ${\bf M}$ which identifies the row states in the $q$ rows.
The two-row states can be constructed as a direct product of the
row states $\ket{\phi_{m_1}}$ and $\ket{\phi_{m_2}}$ as
\begin{equation}
   \Ket{\Phi^2_{M}} \equiv
                        \ket{\phi_{m_1}} \otimes \ket{\phi_{m_2}}
                        : N_{M} \leqslant \Sigma +\Delta.
   \label{clus_2rank}
\end{equation}
Here, ${\bf M} = (m_1, m_2)$  has the row-state quantum numbers of the two
contributing row states, satisfying the constraint that the total number of 
particles in the two-row state can at most be equal to $\Sigma +\Delta$. 
Furthermore, the number of multirow states can be additionally reduced with
the imposition of more complex constraints on $N_M$. Imposing the constraints
in Eqs.~(\ref{ss_bound}), (\ref{row_bound}), and (\ref{cluster_bound}) retains
only the multirow states satisfying these constraints. This reduces the number
of multirow states. Then, like the row-state quantum number $m$, the multirow
state quantum number $M$ of the selected states may not be consecutive
integers. For improved representation of the multirow states, we follow the same
strategy as in a row state. So, let us consider the set formed by the multirow
state quantum numbers in ascending order
$\mathcal{S} = \{ M_I : M_I <M_{I+1},\; \text{and}\; I, I+1 \in \mathcal{S}'\}$.
Here, I is a sequence label, which forms the index set
$\mathcal{S}' = \{ 1, 2, \cdots \beta^{(2)}\}$, and $\beta^{(2)}$ is the
total number of possible two-row states. Each of the two-row states can then
represented as $\Ket{\Phi^2}_I\equiv \Ket{\Phi^2_{M_I}}$. 
Then, by knowing $M_I$, the pair of the constituent row states $(m_1, m_2)$
can be identified. Here, with the constraints it is to be noted that 
$\beta^{(2)} \leqslant (\beta)^2$. The total possible multirow states obtained
with the constraint on $N_M$ are lower in number compared to the possible
combinations of selected row states $\ket{\phi_{m_1}} \otimes \ket{\phi_{m_2}}$.
Appendix \ref{appendA} illustrates, as an example, the construction of possible
two-row state configurations generated from the selected row states. Similarly,
the three-row states can be constructed as a direct product of two-row state
with label $M$ and a row-state with quantum number $m$ as
\begin{equation}
   \Ket{\Phi^3_{M'}} \equiv 
                  \Ket{\Phi^2_{M}} \otimes \ket{\phi_m}
                        : N_{M'} \leqslant \Sigma +\Delta.
\end{equation}
Here, ${\bf M'}$ is a vector containing the information of the row-state
configurations. And the three-row states, $\beta^{(3)}$ in number, can be
similarly identified with an index $I \in \{1,2,\cdots, \beta^{(3)}\}$
and is represented as $\Ket{\Phi^3}_I \equiv \Ket{\Phi^3_{M'}}$. Following
similar steps, states with more number of row states can be constructed.


\subsection{Basis states with constraints}

To generate the basis states for the entire lattice system with the
constraints, we again use the multirow states as the building blocks.
In short, the basis states can be a multirow state or product of two multirow
states with an appropriate number of rows. However, only the basis states
satisfying the constraint in Eq.~(\ref{cluster_bound}) are selected.
Consider the case of $L = 3$ as an example. The basis states which satisfy
the constraint in Eq.~(\ref{cluster_bound}) are to be selected from the set
$\{ \Ket{\Phi^3_{M}} \}$, and the members of this set are generated 
with the constraints in Eqs.~(\ref{ss_bound}), (\ref{row_bound}), and 
(\ref{clus_2rank}). Let 
\begin{equation}
\ket{\Phi^3_{M'}} \equiv \ket{\Phi^3}_I \in \{ \Ket{\Phi^3_{M}} \}
   : \Sigma \leqslant N_{M'} \leqslant \Sigma +\Delta ,
\end{equation}
be one of the three-row states which satisfies all the constraints.
Here $N_{M'}$ represents total number of particles in the three-row state.
Then the collection of these states forms the desired basis set 
$\{ \Ket{\Phi^3_{M'}} \}$.
The labeling of the basis states is done similar to the multirow states as 
described earlier. Thus each basis state is labeled as
$\Ket{\Phi^L}_I$, with index $I \in \{1,2,\cdots \Gamma\}$, where
$\Gamma \leqslant \beta^{(3)}$ is the total number of states in
$\{ \Ket{\Phi^3_{M'}} \}$.
Similarly, for the case of a larger lattice $L =5$, we can use the three- and
two-row states to generate basis states as
\begin{equation}
\!\!\!\!\Ket{\Phi^{5}}_I = \Ket{\Phi^{3}_{M}} \otimes
\Ket{\Phi^{2}_{M'}}:
   \Sigma \leqslant N_M + N_{M'} \leqslant \Sigma +\Delta. 
 \label{basis_stred}
\end{equation}
Once $\{ \Ket{\Phi^5}_I \}$ is generated, the model Hamiltonian 
can then be diagonalized to obtain the ground state as a linear combination,
\begin{equation}
   \ket{\Psi} = \sum_{I=1}^{\Gamma} C^I \, \ket{\Phi^L}_I,
\end{equation}
where $C^I$ is the coefficient corresponding to the contribution of
the basis $\ket{\Phi^L}_I$.


\subsection{Calculation of Hamiltonian matrix elements}

Consider the BHM Hamiltonian in Eq.~({\ref{ham}}), and as an example select 
one of the hopping terms $-J \hat{b}^{\dagger}_{p+1, q}\hat{b}_{p, q}$. Then,
its operation on the row and basis states can be evaluated using the 
expressions in Eq.~(\ref{hop_phi})-(\ref{diff-Phi}). For the $I$th basis state,
$\Ket{\Phi^L}_I \equiv \Ket{\Phi^L_R}$, with the row-state components given by
vector $\bf R$, we get
\begin{equation}
   -J \hat{b}^{\dagger}_{p+1, q}\hat{b}_{p, q} \Ket{\Phi^L_R}
	       = -J \sqrt{n_{p,q}\,(n_{p+1,q} +1)} \Ket{\Phi^L_{R'}}.
\label{ham_mat}
\end{equation}
Here, both states $\Ket{\Phi^L_R}$ and $\Ket{\Phi^L_{R'}}$ are members of
the basis set $\{\Ket{\Phi^L}_{I''} \}$ obtained with the constraints.
From earlier considerations, the row-state quantum numbers of
the two basis states ${\mathbf R} = (r_1, r_2, \cdots r_L)$ and
${\mathbf R'} = (r'_1, r'_2, \cdots r'_L)$ are related by $r'_j = r_j$ for 
$j \ne q$ and $r'_q = r_q + N_B^{p-1} \big( N_B -1 \big)$.
This uniquely identifies ${\mathbf R'}$, but the corresponding 
index $I'$ of $\Ket{\Phi^L_{R'}} \equiv \Ket{\Phi^L}_{I'}$ is yet unknown.
This is due to the lack of an algebraic relation between $I'$ and 
${\mathbf R'}$. Such a relation as in Eq.~(\ref{int_label}) is applicable
when there is no state reduction. Once the indices are known, we can define 
the Hamiltonian matrix 
$H_{I\,I'} = \Bra{\Phi^L_{ R'}} H_{\text{ BHM}}\Ket{\Phi^L_R} $ and 
continue the process for all the hopping terms in the BHM Hamiltonian to 
generate the Hamiltonian matrix.


\subsection{Identifying the basis index $I'$ of ${\bf R}'$}

 For a system with $L$ rows, as mentioned earlier, the $I$th cluster or basis 
state satisfying the constraint equations is 
$\Ket{\Phi^L}_I  \equiv \Ket{\Phi^L_M}$. The corresponding vector label
of the state ${\bf M} = (m_1, m_2, \cdots, m_L)$ has the row-state quantum 
numbers of the constituent rows as components. This forms a forward map 
between $I$ and row-state configurations defined in ${\bf M}$:
\begin{eqnarray}
\Ket{\Phi^L}_I &&\rightarrow \ket{\phi_{m_1}} \otimes
	\ket{\phi_{m_2}} \cdots \otimes \ket{\phi}_{m_L}, \\ \nonumber
		  &&\equiv\; \ket{\phi}_{i_1} \;\otimes\;
	\ket{\phi}_{i_2}\; \cdots \otimes\; \ket{\phi}_{i_L}.
\end{eqnarray}
However, we also require the inverse map. That is, for a given ${\bf M}$ it 
should be possible to identify the corresponding $I$. As discussed earlier,
this mapping information is required in the calculation of the  Hamiltonian 
matrix. Such a mapping can be established using the structure adopted in the 
construction of the cluster states and using bisection. Recall that the number 
of row-state configurations satisfying the constraint is $\beta$, and each 
row state is identified with a quantum number $m$ and a corresponding
label $i$ as $\ket{\phi_m} \equiv \ket{\phi}_i$. The cluster states
$\Ket{\Phi^L}_I$ are identified and sorted in terms of row-state quantum 
numbers $(m_1, m_2, \cdots, m_{L-1}, m_L)$ and  corresponding indices 
$(i_1, i_2, \cdots, i_{L-1}, i_L)$. In this notation, the row-state index 
$i_{q+1}$ varies faster than the $i_q$ with $I$. As an example, the sequence 
of the row-state indices for the cluster states are shown in shown in 
Table~\ref{csi}. From Table~\ref{csi} it is evident that $i_1$ varies 
monotonically with $I$ in the interval $I \in [1,\Gamma]$. However, 
$i_2$ varies monotonically with $I$ for the same $i_1$. Similarly $i_q$
varies monotonically with $I$ in the limited interval where
$(i_1, i_2, \cdots i_{q-1})$ remain unchanged with $I$. Given the sequence, 
for a cluster state with the row-state configurations with indices 
$( i'_1, i'_2, \cdots, i'_L)$, the corresponding cluster state index 
$I'$ can be identified using a set of bisection searches.

\begin{table}[t]
 \begin{tabular}{ |c|c| }
  \hline
      $I$   &$(    i_1,           i_2,      \cdots        i_{L-1},   i_L   )$\\
  \hline                                                          
      $1$   &$(    1,\;\;         1,\;      \;\cdots\;      1,\;\;    1    )$\\
  \hline                                                          
      $2$   &$(    1,\;\;         1,\;      \;\cdots\;      1,\;\;    2    )$\\
  \hline                                                          
   $\vdots$ &$   \vdots\;  \quad \vdots  \qquad\quad  \vdots\quad \;\vdots  $\\
  \hline                                                          
    $I_1$   &$(    1,\;\;         1,\;      \;\cdots\;      1,\;\;  \beta )$\\
  \hline                                                          
  $I_1+1$   &$(    1,\;\;         1,\;      \;\cdots\;      2,\;\;     1   )$\\
  \hline                                                          
   $\vdots$ &$   \vdots\;  \quad \vdots  \qquad\quad  \vdots\quad \;\vdots  $\\
  \hline                                                          
    $I_2$   &$(    1,\;\;         1,\;      \;\cdots\;      2,\;\;  \beta )$\\
  \hline                                                          
   $\vdots$ &$   \vdots\;  \quad \vdots  \qquad\quad  \vdots\quad \;\vdots  $\\
  \hline                                                          
    $I_3$   &$(    1,\;\;         1,\;      \;\cdots\;  \beta,\;\; \beta )$\\
  \hline                                                          
   $\vdots$ &$   \vdots\;  \quad \vdots  \qquad\quad  \vdots\quad \;\vdots  $\\
  \hline                                                          
    $I_4$   &$(    1,\;\;       \beta,\;  \;\cdots\;   \beta,\;\; \beta )$\\
  \hline                                                          
  $I_4+1$   &$(    2,\;\;         1,\;      \;\cdots\;      1,\;\;     1   )$\\
  \hline                                                          
   $\vdots$ &$   \vdots\;  \quad \vdots  \qquad\quad  \vdots\quad \;\vdots  $\\
  \hline                                                          
    $I_5$   &$(    2,\;\;       \beta,\;  \;\cdots\;   \beta,\;\; \beta )$\\
  \hline                                                          
   $\vdots$ &$   \vdots\;  \quad \vdots  \qquad\quad  \vdots\quad \;\vdots  $\\
  \hline                                                          
   $\Gamma$  &$( \beta,\;\;     \beta,\;  \;\cdots\;   \beta,\;\; \beta )$\\
  \hline
 \end{tabular}
\caption{Table showing the row-state indices $i_q$ of the row-state
	 configurations in all the rows that together constitute the $I$th
	 basis state.}
\label{csi}
\end{table}

Let us suppose that we have a basis state with row-state configuration
${\bf R}'=( r'_1, r'_2, \cdots, r'_L)$ and labels $( i'_1, i'_2, \cdots,i'_L)$,
and we have to identify the corresponding
state index $I'$ from the basis set. To accomplish this, we perform a search 
using the method of bisection in the interval $I \in [1, \Gamma]$. Considering 
that the leftmost index $i_1$ as the slowest varying index, we start the
bisection to identify the range of basis set index
$I\in [I_{\rm min}^1, I_{\rm max}^1]$, which has $i'_1$ as the leftmost
row-state index within the basis set. Next, a bisection search is done within
this range to locate the range $ [I_{\rm min}^2, I_{\rm max}^2]$ which 
has the row-state index $i'_2$. At this stage all the basis states within 
the range $ [I_{\rm min}^2, I_{\rm max}^2]$ have the same  $i'_1$ and
$i'_2$ as the desired state ${\bf R}'$. We continue the bisection further
for each of the row-state indices, and the required basis-state index $I'$ is
obtained in the last bisection with the identification of the $i'_L$.
Thus the inverse map needed for locating the basis index from row-state
configurations can be achieved through a sequence of $L$ bisection searches.
It should be noted that each bisection search requires a maximum of
$\log_2(\Gamma)$ comparisons of the row-state index.
This is a huge simplification compared to the $\Gamma$ number of
searches required in locating the desired ket with nonzero matrix element, in
the brute-force way of spanning over all possible basis states.

\begin{figure}[t]
 \begin{center}
   \includegraphics[width=8.5cm]{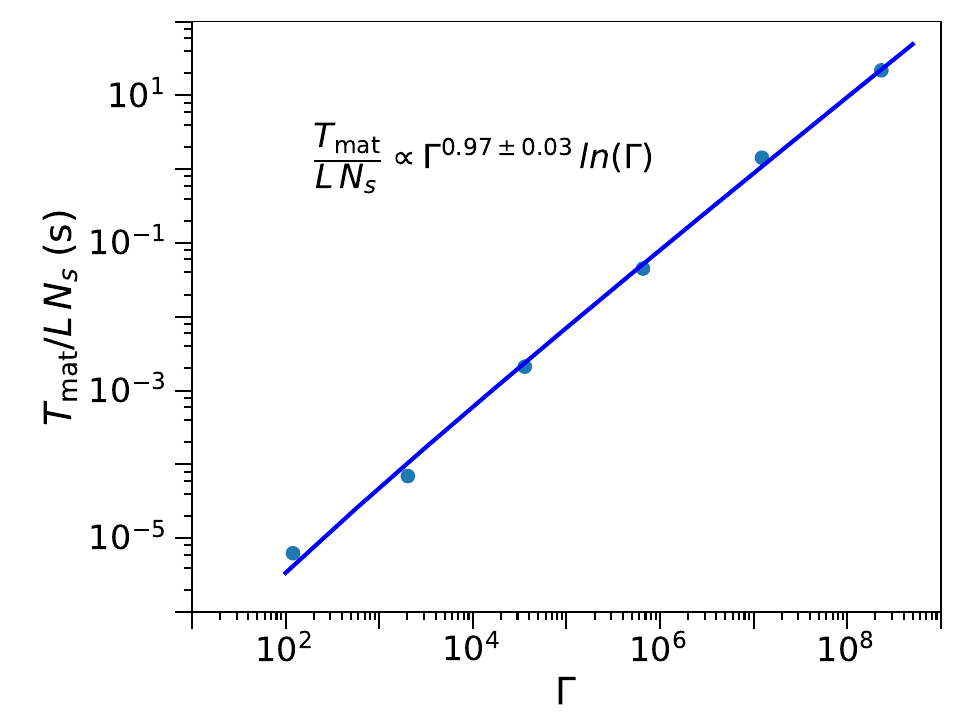}
    \caption{Scaling of the matrix computation time ($T_{\rm mat}$) with the
             size of the basis set $\Gamma$. The filled circles
             corresponds to the $T_{\rm mat}$ of our code, and the solid line
             is the fitted curve.
             }
  \label{mat_time}
 \end{center}
\end{figure}

Using this method we can, then, generate the Hamiltonian matrix similar to
the approach described in Sec. \ref{sect-ham-mat}. For a
state  $\Ket{\Phi^L}_I$, we identify the $_{I'}\Bra{\Phi^L}$ which has 
nonzero matrix element for each of the terms in the Hamiltonian and calculate
the matrix element. This generates the Hamiltonian matrix elements along
the column corresponding to $\Ket{\Phi^L}_I$. Continuing the process for 
other states, the Hamiltonian matrix can be calculated column wise.
It is expected that the computational time $T_{\rm mat}$
involved in the matrix construction scales with the total number of hopping
terms in the Hamiltonian, which is proportional to the number of lattice sites
$N_s$. And as mentioned earlier, for each ket state the bra state with a
nonzero matrix element is identified in $\sim L\times \ln(\Gamma)$ 
bisection steps. Thus, it is expected that $T_{\rm mat}/N_{s}$ should scale as
$\sim \Gamma L\,\ln(\Gamma)$. This dependence is demonstrated in 
Fig.~\ref{mat_time}, for the BHM Hamiltonian with hardcore bosons on a 2D 
lattice of dimension $8\times L$, with $L \in \{2,3,4,5,6,7 \}$ and at fixed 
number density $N/N_s = 1/8$, where different lattice sizes determines the 
$\Gamma$. From the figure it can be noted that
$T_{\rm mat}/N_s \propto \Gamma^{0.97 \pm 0.03}\,L\,\ln(\Gamma)$, which is 
consistent with the scaling relation obtained earlier.

Using the method described, we can calculate the Hamiltonian matrix and
diagonalize it to obtain the eigenspectrum and states of the system.
To demonstrate the reliability of our ED method, we discuss 
selected results of BHM computed using the method in the next section.


\section{ED results for ground state of BHM} \label{sec_bhm_res}
In this section we discuss selected ground-state properties of BHM, computed
using our ED method. First, we demonstrate the accuracy of results, and
later we discuss the advantage of our method. We highlight the computational
advantages of the state reduction.


\subsection{Ground state of BHM at unit filling}
\begin{figure}[t]
 \begin{center}
   \includegraphics[width=8.5cm]{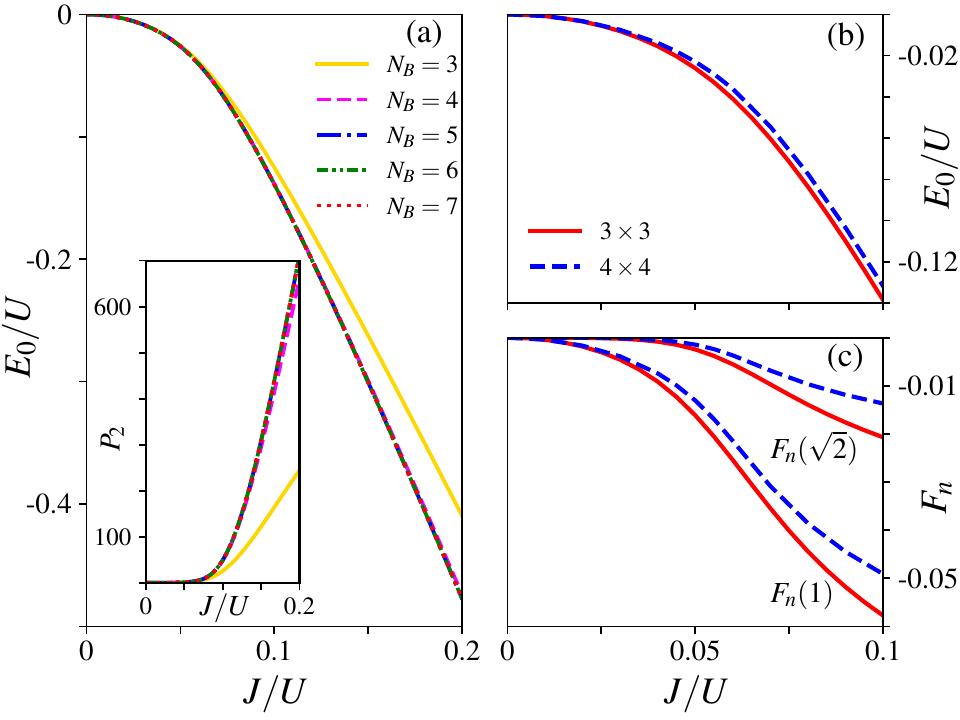}
	 \caption{Ground-state properties of BHM with unit-filling.
                  Panel (a) shows the effect of varying the cutoff in the
                  single-site Fock-space basis size $N_B$ on the energy, 
                  and the inset plot corresponds to IPR for a $3 \times 3$
                  system. These quantities are robust for $N_B \gtrapprox 4$,
                  as seen from the curves sitting on top of each 
                  other for NB = 4 and higher.
                  Panels (b) and (c) show the energy per site $E_0/U$ and the
                  particle-number correlation functions $F_n(r)$, respectively,
                  computed for $3\times 3$ (solid lines) and $4 \times 4$ 
                  lattices (dashed lines).}
  \label{gs_bhm}
 \end{center}
\end{figure}

In the calculations with softcore bosons, it is essential to introduce a
cutoff to the single-site Fock space $N_B$. This restricts the maximum
occupancy of a site. Here, we show that an appropriate $N_B$ can
be chosen without affecting the accuracy of the ground state. We demonstrate
this by varying $N_B$ and computing the ground-state energy for a $3\times 3$
system with unit average occupancy. As shown in Fig.\ref{gs_bhm}(a), the
ground state-energy per lattice site $E_0$ remains unchanged with
$N_B \gtrapprox 4$. Another property of interest is the inverse participation
ratio (IPR) $P_2=\sum_I |C^I|^4$, which measures the spread of
ground state $\Ket{\Psi} = \sum_I C^I \Ket{\Phi}_I$ among the basis states.
As seen from the inset of Fig. \ref{gs_bhm}(a), $P_2$ remains
the same for $N_B \gtrapprox 4$, which suggests it as good choice for the
cutoff. Further, for a $4 \times 4$ system with $J/U = 0.1$ and unit
occupancy, the ground-state energy per lattice site is
$E_0/U = -0.1176$, $-0.1318$, and $-0.1321$ for $N_B = 3$, $4$, and $5$,
respectively. Here, the change in the energy between $N_B = 4$ and $5$ is
negligible; thus $N_B \gtrapprox 4$ is an appropriate cutoff for
$J/U \lesssim 0.1$.

To benchmark our results, we choose $N_B =4$ for computing the ground-state 
energy of a $3 \times 3$ lattice with unit filling as a function of $J/U$. The 
results are shown in Fig.\ref{gs_bhm}(b) which are in excellent agreement 
with the ED results of the same system size reported in 
Ref. \cite{krutitsky_16}. Furthermore, by increasing the system size to 
$4 \times 4$, we note that the computed energy approaches the QMC results. 
We also compute the particle-number correlation function in the ground state 
of the BHM,
\begin{equation}
  F_n(r) = \langle\hat{n}_{\bf r}\hat{n}_{\bf 0}\rangle 
 - \langle\hat{n}_{\bf r} \rangle \langle\hat{n}_{\bf 0} \rangle,
\end{equation}
which measures the density-density correlations between two lattice
sites, one of which is fixed at the origin and the other site is located at
the coordinate ${\bf r}$. Figure \ref{gs_bhm}(c) shows the variation of 
$F_n(r)$ for $r =|{\bf r} - {\bf 0}| = 1$ and $\sqrt{2}$ as a function
of $J/U$. Here, $r=1$ and $\sqrt{2}$ correspond to the density 
correlations between sites which are nearest-neighbors and next-nearest 
neighbors, respectively. The results of the $3 \times 3$ lattice 
agree with the ED results of the same system size reported in 
Ref. \cite{krutitsky_16}. And the results of the $4 \times 4$ system are close 
to the QMC data. For further comparisons with QMC results, we compute 
the ground-state phase diagram of BHM. 

\subsection{Phase diagram of BHM}
\begin{figure}[t]
 \begin{center}
   \includegraphics[width=8.5cm]{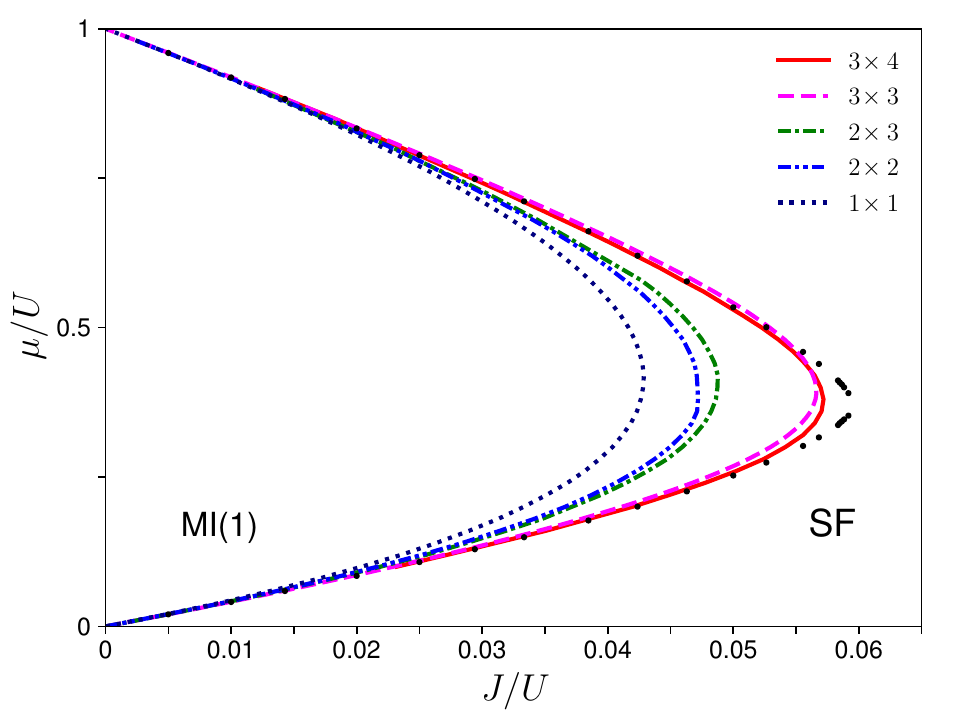}
    \caption{Phase diagram of BHM computed using CGMF with varying cluster 
             sizes by employing periodic boundary conditions along $x$ and 
             $y$ directions via mean field. The plot corresponding to 
             $1\times 1$ represents data obtained from the single-site 
             mean-field method.
             For $3\times 3$ and $3\times 4$ clusters, the periodic boundary
             condition along the $x$ direction is considered as exact hopping.
             With larger clusters, the Mott lobe approaches the QMC
             results from Ref.~\cite{cappogrosso_08}, shown with black filled
             circles \cite{qmc_comm}.}
  \label{bhm_phdiag}
 \end{center}
\end{figure}

To check the reliability of the method, we next calculate the phase diagram
of BHM using the cluster Gutzwiller mean-field (CGMF) method. 
In the CGMF method, the lattice is tiled with clusters, which are coupled
through the mean-field, and the wave function of each cluster is calculated
using the exact diagonalization. Thus the diagonalization of a Hamiltonian 
matrix corresponding to a cluster forms the core of the CGMF method. 
Consequently, calculation of the phase diagram and comparing with standard 
results from QMC serves as an important means to check the results obtained 
using our ED method.  More details on the CGMF method are given 
in Refs.\cite{luhmann_13, bai_18}. After rescaling the BHM Hamiltonian with 
the on-site interaction strength $U$, we generate the ground-state phase 
diagram in the $\mu/U$ vs $J/U$ plane as shown in Fig.~\ref{bhm_phdiag}.
Here, $\mu$ is the chemical potential that fixes the total number of 
particles in the system. As the size of the cluster is increased, the tip of 
a Mott lobe, which defines the phase boundary between the Mott insulator (MI) 
and superfluid (SF) phases, shifts to higher $J/U$. With single-site 
mean-field method (labeled as $1 \times 1$), the tip of the Mott lobe lies 
at $J_c/U = 0.0429$. And, with CGMF using a $3 \times 4$ cluster, the tip 
shifts to $0.0572$.  A cluster-size scaling analysis reported in 
Ref.~\cite{bai_thesis} gives the location of the tip as $J_c/U = 0.0595$, which 
is very close to the QMC result of $0.0597$ reported in 
Ref.~\cite{cappogrosso_08}. Thus, the consistency of the CGMF result with
the QMC result directly establishes the accuracy of our method.

\subsection{Computational advantage with our ED method}
\begin{figure}[t]
   \includegraphics[width=8.5cm]{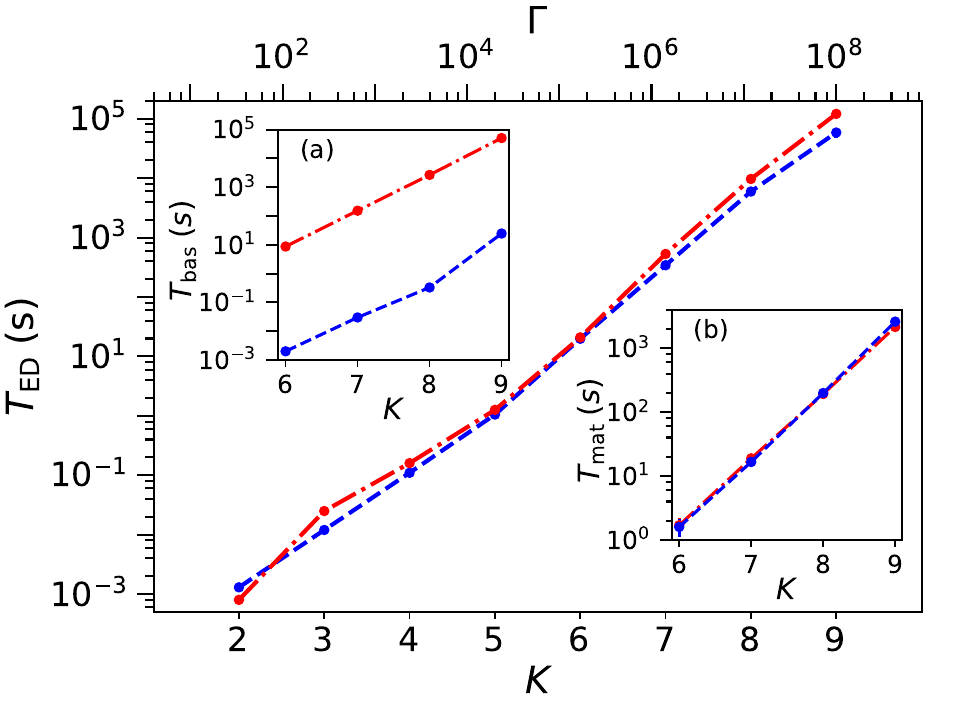}
	\caption{Comparison of the results from our ED code (dotted curves)
                 with an open-source ED package \cite{zhentao_wang}
                 (dash-dotted curves). The elapsed 
                 time in computing the ground state is plotted for a 
                 system of hardcore bosons at density $1/4$ on a $K \times 4$ 
                 lattice with $\Gamma$ as total basis states. Inset (a) shows
		 the time taken in the basis construction ($T_{\rm bas}$), and
                 inset (b) shows the time taken in construction of the
                 Hamiltonian matrix ($T_{\rm mat}$). 
                 In inset (b), the two curves sit on top of each other.
                }
  \label{comparison_ED}
\end{figure}

To benchmark the results of our ED method based on hierarchical wave functions, 
we compare it with a standard ED  library named ``quantum basis" available at 
Github~\cite{zhentao_wang}. Using the two codes we check the elapsed time in 
ED $(T_{\rm ED})$ as a function of the lattice size and 
corresponding basis set dimension $\Gamma$. For this we chose a
system of hardcore bosons at fixed density of $1/4$, for different system
sizes $2\times 4$, $3\times 4$, $\cdots$, and  $9\times 4$.  Additional 
details of the study are given in Appendix \ref{appendB}. The plot 
comparing the performance of the two codes is shown in 
Fig.~\ref{comparison_ED}. It shows that our code performs better.
The basis construction procedure using hierarchical wave functions offers 
advantages and is better, particularly for larger lattice dimensions. This can 
be seen from the inset (a) of Fig.~\ref{comparison_ED}. For the Hamiltonian 
matrix construction, both the codes take almost the same duration, which
can be seen from the inset (b) of Fig.~\ref{comparison_ED}. Similarly, matrix 
diagonalization takes similar time. This is expected, as both the codes use 
{\sc ARPACK} for matrix diagonalization.

\begin{figure}[t]
   \includegraphics[width=8.5cm]{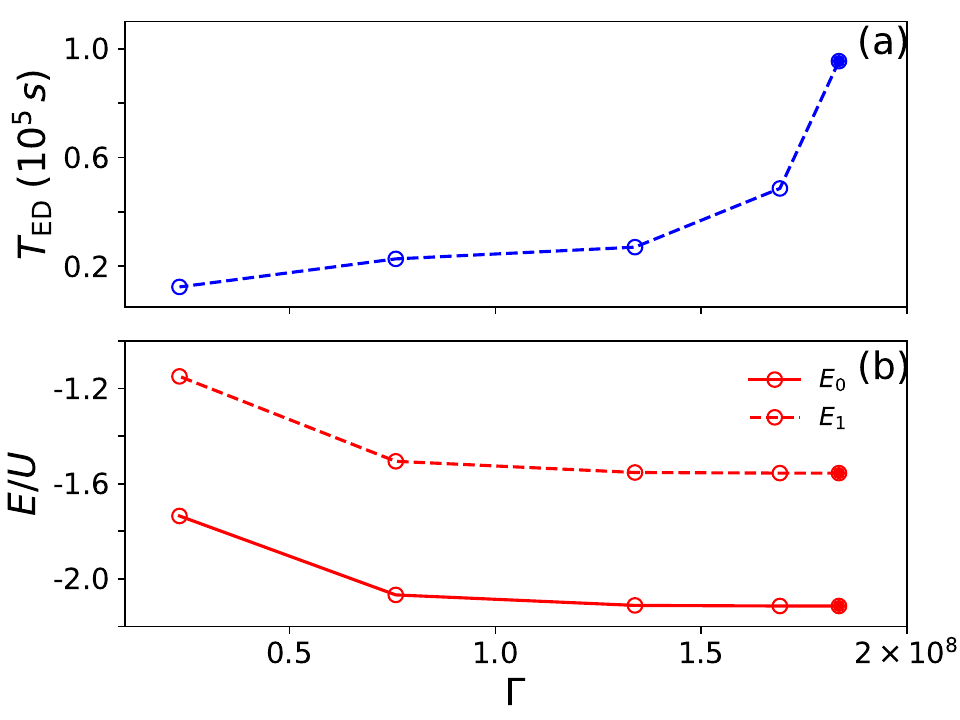}
	\caption{Comparison of the ED results against the basis set dimensions
                 based on a set of state reductions. For the ground state
                 of a $4\times 4$ system with unit filling, the system 
                 parameters are chosen as $N_B =5$ and $J =0.1U$. The state 
                 reduction constrains the basis states by allowing the total 
                 number of bosons in a row to be four with a variation of  
                 $\pm 1, \pm 2, \pm 3$, and $\pm 4$. Open (filled) circles 
                 represents the results with (without) state reduction, and 
                 solid/dashed lines are visual guide to the eye. Panels (a) 
                 and (b) show the elapsed time in ED, and the energies of the 
                 ground state ($E_0$) and first excited state ($E_1$), 
                 respectively.}
  \label{time_stred}
\end{figure}

To demonstrate the computational advantages with the state reduction, we
compute $T_{\rm ED}$ of $4\times 4$ system at unit filling
with $N_B =5$ and $J = 0.1U$. In addition, the accuracy of the results with
state reduction can be checked by changing the constraints and
noting the variations in results. For this, we choose a set of state 
reductions, introduced in Eq.~(\ref{row_bound}), to constrain 
the number of bosons of each row by 4 with fluctuations $\pm 1$, $\pm 2$, 
$\pm 3$, and $\pm 4$. The trend in $T_{\rm ED}$ with different fluctuations 
is shown in Fig.~\ref{time_stred} (a). The corresponding energies of the 
ground state ($E_0$) and 
first excited state ($E_1$) are shown in Fig.~\ref{time_stred} (b).
The figure shows that significant reduction of $T_{\rm ED}$ 
is achievable without compromising the accuracy of the results. 

In the next section we discuss how our ED method can be employed 
to trace over the spatial degrees of freedom of a wave function when the system 
is bipartitioned. This can be used in characterizing the entanglement of 
a state.



\section{Entanglement in the quantum phase} \label{sec_entang}


The topological characteristics of a system can be inferred by studying the
entanglement in the system. It can be used to distinguish a topologically
ordered phase like quantum Hall from a normal phase like superfluid.
Among various indicators of entanglement in a system, the bipartite 
entanglement entropy is considered robust. In real-space, it is calculated by 
partitioning the lattice system into two subsystems, say A and B.
The many-body ground-state wave function of the system can then be written in
a Schmidt decomposed form as \cite{ekert_95}
\begin{equation}
 \Ket{\Psi} = \sum_{j} \sqrt{\lambda_{j}} \Ket{\Psi_{\rm A}} \otimes
              \Ket{\Psi_{\rm B}}.
\end{equation}
Here, $\lambda_{j}$ are the Schmidt coefficients, and these are the
eigenvalues of the reduced density matrix of the subsystem. The reduced 
density matrix associated with subsystem A can be obtained from the density 
matrix after tracing over the degrees of freedom associated with the 
subsystem B, $\rho_{\rm A} = Tr_{\rm B} \Ket{\Psi}\Bra{\Psi}$. The eigenvalues 
$\lambda_i$ of the  reduced density matrix give the entanglement spectra, and 
the bipartite entanglement entropy of the system is the von Neumann entropy 
associated with the reduced density matrix,  
\begin{equation}
   S_{E} = -\text{Tr}[\rho_{\rm A} \ln{\rho_{\rm A}}] = 
                           -\sum_{j} \lambda_{j} \ln{\lambda_{j}}.
\end{equation}
The entanglement entropy $S_{E}$ scales with the length $(L)$ of the interface 
between the two subsystems as \cite{kitaev_06, levin_06} 
\begin{equation}
  S_{E} = \beta L -\gamma + \mathcal{O}(L^{-\nu})\;,\; \nu > 0,
  \label{s_scal}
\end{equation}
where $-\gamma$ is the topological entanglement entropy. It measures the 
quantum dimension of the quasiparticle excitations in the FQH states.
The topologically ordered $\nu=1/m$ FQH state has $\gamma = 1/2\ln(m)$.
The area-law scaling of the entanglement is typical of ground states for gapped
systems \cite{eisert_10}. A different way of characterizing the entanglement
is to use the geometric entanglement, which does not involves the
bipartitioning of the lattice \cite{zhang_17}.


\subsection{Constructing bipartite reduced density matrix}

\begin{figure}[t]
 \begin{center}
   \includegraphics[width=8.5cm]{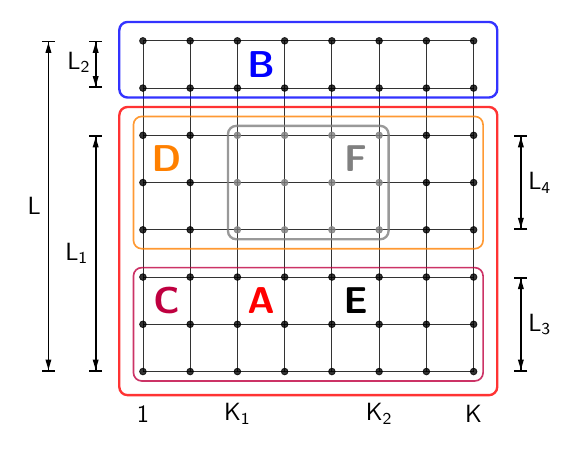}
    \caption{Visual illustration for the bipartition of the lattice
             for obtaining the reduced density matrix of subsystem F.
             First, the lattice is partitioned along the $y$ axis into
             subsystems A and B. Then subsystem A is further subdivided
             into subsystems C and D. Finally, subsystem D is partitioned
             along the $x$ axis to give subsystem F,
             which comprises the gray-colored lattice sites.
             The boundaries of the subsystems are identifiable from
             the corresponding color or shading of the subsystem label.
             The subsystem E corresponding to
             the surrounding of subsystem F constitutes the black-colored
             lattice sites.}
  \label{bipartition}
 \end{center}
\end{figure}

The approach adopted in the construction of the basis set can be employed
in an optimal way to calculate the bipartite reduced density matrix. For this
consider partitioning a lattice system with $L$ rows into two subsystems A
and B such that the subsystem A consists of bottom $L_1$ rows and  B 
consists of top $L_2 = L - L_1$ rows. This is shown schematically in 
Fig.~\ref{bipartition}. As each basis state of the lattice consists of $L$
row states, and following Eq.~(\ref{basis_stred}), we can write
\begin{equation}
 \Ket{\Phi^L_{M}} \equiv \Ket{\Phi^{L_1}_{M_1}} \otimes
                \Ket{\Phi^{L_2}_{M_2}}:
                \Sigma \leqslant N_{M_1} + N_{M_2} \leqslant
                \Sigma +\Delta. 
\end{equation}
Each basis state $\Ket{\Phi^L_{M}}$, labeled with index $I$, is thus 
the direct product of multirow states $\Ket{\Phi^{L_1}_{M_1}}$ 
and $\Ket{\Phi^{L_2}_{M_2}}$  with labels $I_1$ and $I_2$, respectively.
The reduced density matrix $\rho_{\rm A}$ would then be a 
$\beta^{(L_1)}$-dimensional matrix and is obtained after tracing over the 
degrees of freedom associated with $I_2$:
\begin{eqnarray}
 \rho_{\rm A}(k,l) &=& \sum_{I=1}^{\Gamma} \sum_{I'=1}^{\Gamma}
                 C^{*I} \, C^{I'}\,
                 \delta_{I_2 \,, I'_2} \;
                 \delta_{I_1 \,, k} \; \delta_{I'_1 \,, l}
                 \nonumber \\
                 &=& \sum_{I=1}^{\Gamma} \sum_{I'=1}^{\Gamma}
                 \rho(I,I') \,
                 \delta_{I_2 \,, I'_2} \;
                 \delta_{I_1 \,, k} \; \delta_{I'_1 \,, l} .
\end{eqnarray}
Here $\rho = |\Psi \rangle\langle \Psi | $ is the density matrix of the system.
Similarly, the subsystem A can be further partitioned into two
subsystems C and D consisting of bottom $L_3$ rows and top
$L_4 = L_1 - L_3$ rows, respectively. Thus the subsystem state
$\Ket{\Phi^{L_1}_{M}}$, labeled with index $I$ for 
$1 \leqslant I \leqslant \beta^{(L_1)}$, is the direct product of
the state of subsystems C $\Ket{\Phi^{L_3}}_{M_3}$ and D 
$\Ket{\Phi^{L_4}}_{\bf M_4}$ with corresponding
indices $1 \leqslant I_3 \leqslant \beta^{(L_3)}$ 
and $1 \leqslant I_4 \leqslant \beta^{(L_4)}$, respectively. Thus 
the $\beta^{(L_4)}$-dimensional reduced density matrix $\rho_{\rm D}$ can be 
obtained as
\begin{equation}
 \rho_{\rm D}(k,l) = \sum_{I =1}^{\beta^{(L_1)}}
               \sum_{I' =1}^{\beta^{(L_1)}}
	       \rho_{\rm A}(I,I') \, \delta_{I_3 \,, I'_3} \;
               \delta_{I_4 \,, k} \; \delta_{I'_4 \,, l} \,.
\end{equation}

The bipartitioning of the system discussed so far is by separating along the 
rows. An example of the partition geometry is as indicated by the sublattices 
A and B in Fig.~\ref{bipartition}.  However, following ref. \cite{kitaev_06},
the calculation of entanglement entropy requires bipartitioning the system
into an isolated and surrounding domains. This is as indicated by the 
subsystem F and the remaining part E in Fig.~\ref{bipartition}.  In the figure,
the gray-colored lattice sites constitute the subsystem F, and 
the rest of the lattice sites in black constitute the subsystem E. The 
boundary enclosing the gray-colored lattice sites
marks the bipartite separation between subsystems F and E.  In this case, 
the partitioning involves both along rows and columns. To describe the 
partitioning with a specific example, consider the subdivision of the whole
system in Fig.~\ref{bipartition} into two, the isolated region F and the 
remaining part E. To calculate the reduced density matrix of the subsystem F, 
first the subsystems B and C can be traced out using the method discussed 
earlier. Then, the required reduced density matrix is obtained by partitioning
subsystem D into subsystem F and the remaining lattice sites which forms 
a part of the subsystem E as shown in Fig.~\ref{bipartition}.
The surrounding of subsystem F corresponds to black-colored lattice sites 
which constitutes the subsystem E. The partitioning involves
tracing out columns on the left and right of subsystem F. Thus, the lattice 
sites $(p,q)$ for which $ L_3+1 \leqslant q \leqslant L_1$ and
$p \in \{ K_1, K_1 + 1, \cdots p, \cdots K_2 \}$ lie in subsystem $F$ and the 
rest of the sites are a part of subsystem E over which we wish to trace out.
This necessitates the introduction of a unique label corresponding to the 
different configuration of particles in subsystem F. If the total number of 
sites in the subsystem F is not large, we can use the following scheme for 
assigning each configuration with a unique label $\tilde{m}$ given by 
\begin{equation}
\tilde{m} = \sum_{q=L_3 + 1}^{L_1} \sum_{p=K_1}^{K_2}
                n_{p,q} \,
		(N_B)^{(p-K_1) + (L_1-q)(K_2-K_1+1)}.
        \label{map3}
\end{equation} 
For each multirow state $I$ of subsystem D, the configuration of particles
in the subsystem F corresponds to label $\tilde{m}$ given by Eq.~(\ref{map3}).
For large $N_B$, the label $\tilde{m}$ can take very large 
values. This can be circumvented by using the minimal single-site occupancy
$\eta$ as a shift in the occupancies $n_{p,q} \rightarrow n_{p,q} -\eta$ and
$N_B \rightarrow N_B -\eta$. 
Thus, the reduced density matrix of subsystem F can be expressed as
\begin{equation}
 \rho_{\rm F}(k,l) = \!\sum_{I =1}^{\beta^{(L_4)}}
	     \sum_{I'=1}^{\beta^{(L_4)}} \rho_{\rm D}(I,I')
	     \delta_{\tilde{m}, k} \delta_{\tilde{m}', l} \prod_{p,q}^{'}
             \delta_{n_{p,q}, n'_{p,q}},
        \label{rhoF}
\end{equation}
where $n_{p,q}$ and $n'_{p,q}$ are occupancies at the lattice site $(p,q)$ 
corresponding to the multirow states with indices $I$ and $I'$, respectively.
And the prime on the product over lattice sites $(p,q)$ signifies that
product is restricted over only those lattice sites for which 
$p \in \{1,2,\cdots K_1-1, K_2+1,\cdots K \}$ and
$q \in \{L_3 +1, L_3 +2, \cdots L_1 \}$. Alternatively, if the total number of
lattice sites in subsystem F is large, then instead of assigning a label to
each configuration using Eq.~(\ref{map3}), we can construct the multirow
states using row states of $K_2 - K_1$ lattice site. And each configuration of
particles in subsystem F can then be uniquely identified with the multirow
index described earlier. It should be emphasized that the strategy described
for obtaining $\rho_{\rm F}$ is computationally advantageous, since tracing out 
entire rows can be easily done by utilizing the hierarchical wave functions.


\subsection{Fractional quantum Hall on a lattice}

\begin{figure}[t]
 \begin{center}
   \includegraphics[width=8.5cm]{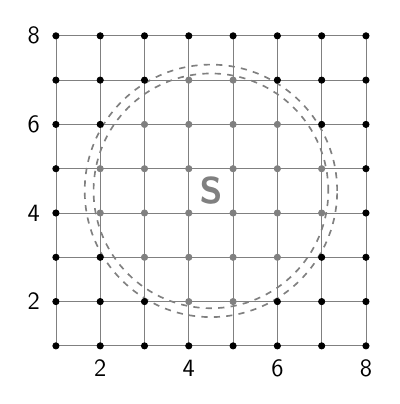}
    \caption{Visual illustration for spatial bipartitioning of the lattice
             such that the lattice sites within the circle (gray-colored
             sites) are chosen as a part of the subsystem $S$.}
  \label{bipart_circles}
 \end{center}
\end{figure}

\begin{figure}[t]
 \begin{center}
   \includegraphics[width=8.5cm]{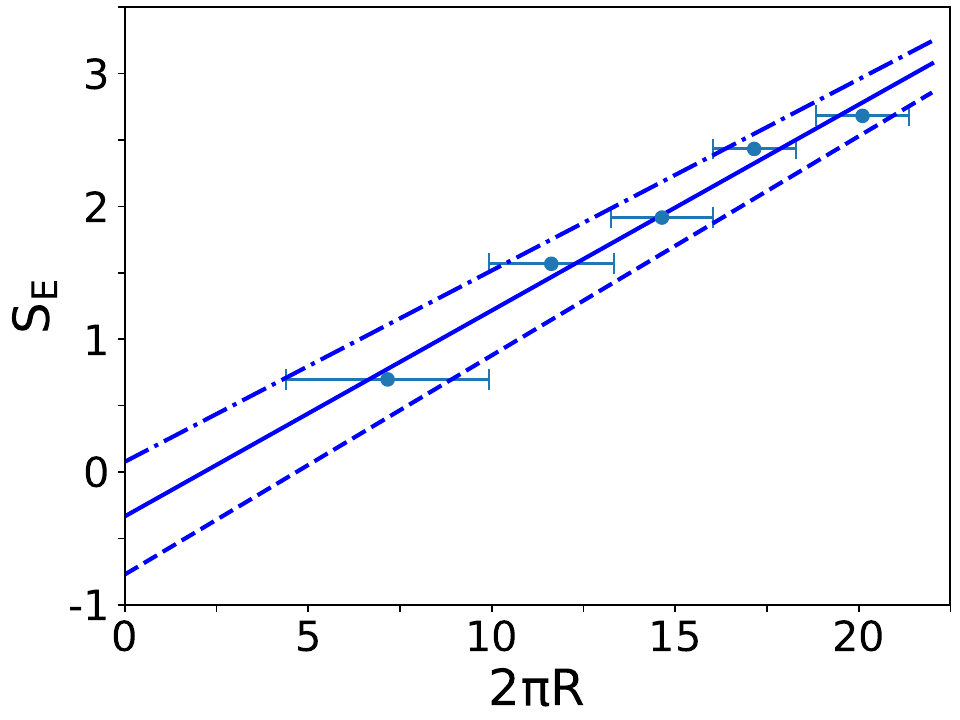}
    \caption{Entanglement entropy $(S_E)$ of the subsystem $S$ as a function of
             the length $2\pi R$ of the circular boundary separating $S$ from 
	     the rest of the lattice. The plot shows $S_E$ for different 
             subsystem boundary $S$, and the $x$ error bars represent the
             ambiguity in the definition of boundary length. The solid, dashed,
             and dash-dotted lines are the curve obtained by fitting 
             the average, minimum, and maximum values for the boundary length,
             respectively.}
  \label{entropy_circles}
 \end{center}
\end{figure}

The topological entanglement entropy $\gamma$ is an appropriate property to 
distinguish quantum Hall states from the other competing states which are 
not topological in nature. As an application of the scheme we have developed, 
we consider the FQH state on a square lattice. The FQH states are supported in
the optical lattice with the introduction of synthetic magnetic fields. With 
this, the hopping term of BHM acquires a lattice-site-dependent Peierls phase.
The Hamiltonian of the system in the Landau gauge is given by \cite{jaksch_03}
\begin{eqnarray}
   \hat{H} = \sum_{p, q}\bigg [ &&\Big( -J e^{i 2\pi \varphi}
             \hat{b}^{\dagger}_{p+1, q}\hat{b}_{p, q} 
             -J \hat{b}^{\dagger}_{p, q+1}\hat{b}_{p, q} + {\rm H.c.}\Big)
             \nonumber\\ 
             &&+ \frac{U}{2} \hat{n}_{p, q} (\hat{n}_{p, q}-1) - 
             \mu\hat{n}_{p, q}\bigg],
\label{ham_mag}  
\end{eqnarray}
where $\varphi$ corresponds to the strength of the magnetic field.
We obtain the $\nu=1/2$ FQH state for a system of four hardcore bosons on a
$8 \times 8$ lattice with periodic boundary conditions and synthetic
magnetic field $\varphi = 1/8$. We find the ground state is doubly degenerate,
which is an essential property of the $\nu = 1/2$ FQH state on a torus 
geometry \cite{wen_90}. Using the prescription of Hatsugai \cite{hatsugai_05},
we have verified the topological nature of the state and find the many-body 
Chern number is 1. To calculate $\gamma$, we bipartition the lattice system as
schematically shown in Fig.~\ref{bipart_circles}. Then, we calculate 
$\gamma$ between the subsystem $S$ and the rest of the lattice. However, as 
the lattice is discrete in nature, bipartitioning it with circular geometry 
introduces an ambiguity in the definition of the boundary as shown in 
Fig.~\ref{bipart_circles}. So, for a given configuration of lattice sites in 
the subsystem $S$, we can consider the boundary of the circle enclosing $S$ 
to lie within the minimum and maximum radii $(R_{\rm min}, R_{\rm max})$. The 
scaling of $S_{E}$ with the boundary length $L$, as shown in 
Fig.~\ref{entropy_circles}, are bounded by two lines, the lower and upper
lines corresponding to least-squares fit of data for circles with radii
$R_{\rm min}$ and  $R_{\rm max}$, respectively. In the figure, the least-squares
fit with the average of the two radii is represented by the middle line. 
We observe that $\gamma$, which is the $y$-intercept, depends on the chosen
definition of the boundary length and varies from $0.08$ to $-0.77$.
Considering the average value of the radius $(R_{\rm min} + R_{\rm max})/2$, 
and a least-squares fit, shown as the middle line in 
Fig.~\ref{entropy_circles}  gives $\gamma = 0.33 \pm 0.17$. This value is 
consistent with the theoretical estimate of 
$\gamma = \ln\sqrt{2} \approx 0.347$ for a $\nu =1/2$ Abelian FQH state.
The numerical estimates can be improved by incorporating entropy calculations
with larger values of $L$ which needs larger lattice dimensions;
however, the ambiguity in the definition of $L$ would still be present.

\begin{figure}[t]
 \begin{center}
   \includegraphics[width=8.5cm]{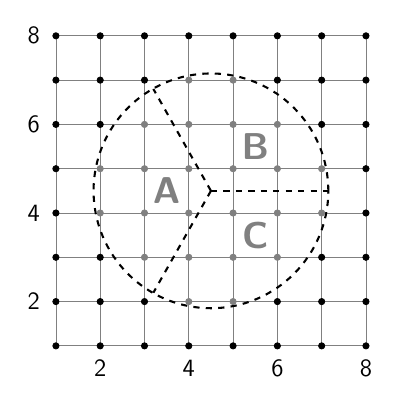}
    \caption{Visual illustration for spatial bipartitioning of the subsystem
             in three slices, forming the plural areas. The gray-colored 
             lattice sites, within the dashed circle, are categorized  
             into subsystems A, B, and C.}
  \label{plural_circles}
 \end{center}
\end{figure}

\begin{figure}[t]
 \begin{center}
   \includegraphics[width=8.5cm]{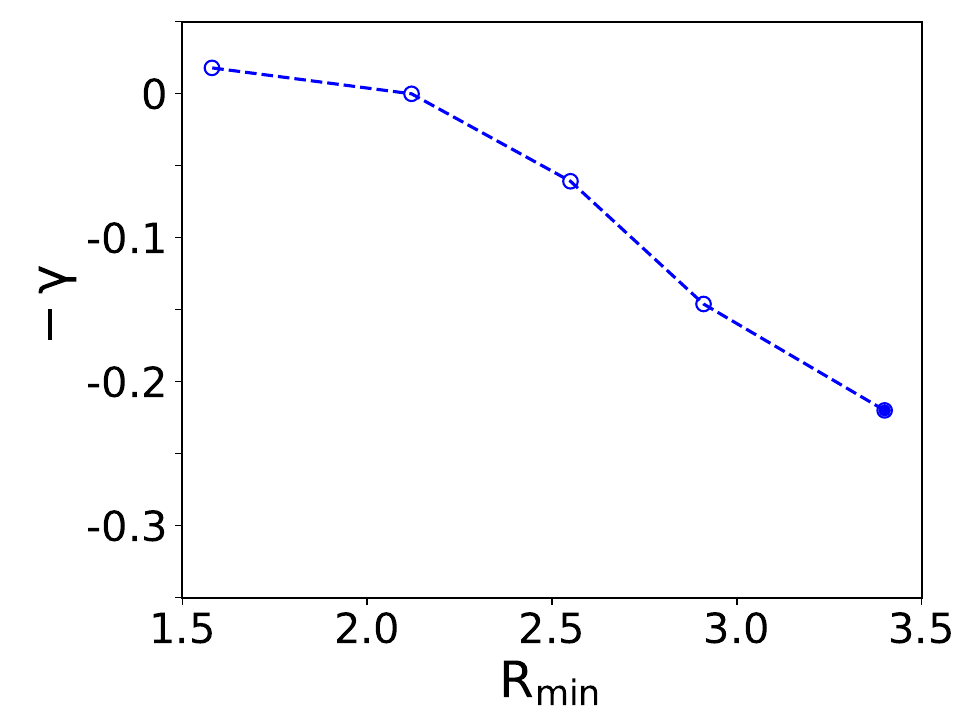}
    \caption{Topological entanglement entropy $(-\gamma)$ obtained for various
             choices of the circular subsystem $S$ with radii $R_{\rm min}$.
             Open circles correspond to the subsystem S, with the center
             located at the center of the $8\times 8$ lattice system,
             and the dashed line is visual guide to the eye. The filled
             circle corresponds to subsystem S, with the center shifted from 
             the system center.}
  \label{tee_circles}
 \end{center}
\end{figure}

Using the prescription of Kitaev and Preskill for the calculation of $\gamma$ 
in terms of the multiple areas, it can be calculated without the ambiguity of
boundary length. In their method $\gamma$ is given by an appropriate
combination of the entanglement entropies of the multiple areas of the
subsystem $S$ (shown in Fig.~\ref{plural_circles}) according to the relation
$S_{ABC} -S_{AB} -S_{BC} -S_{AC} + S_A +S_B + S_C = -\gamma$ \cite{kitaev_06}.
With this formulation we have calculated $\gamma$ for various sizes
of subsystem $S$, and the results are shown in Fig.~\ref{tee_circles}.
We observe that the numerical estimates of $\gamma$ are different from the 
theoretical estimate of $-0.35$. However, the trend of $\gamma$ indicates
approaching the theoretical estimate with the larger dimensions of the
subsystem $S$. This can be due to the requirement of a smooth large-sized
boundary compared to the correlation length $\xi \ll R$. The estimates of
$\gamma$ can be improved with a larger-sized subsystem on a larger lattice,
but the computational resources required is beyond the scope of available
computational resources.


\section{Discussions} \label{sec_discussions}
We report the development of a new method on exact diagonalization. It 
relies on a hierarchy of states to define the basis states of a lattice 
system. The starting point is the Fock states of the single-lattice sites.
Using these, row states are defined as a direct product and then, the product
of the row states forms the multirow states. The approach is flexible and 
it is particularly well suited for studying bosonic optical lattice systems. 
For such a system, there is no limit on the lattice site occupancy, and
the number of basis states grows very fast with the system size and number
of particles. In this multistep and hierarchical approach, we can 
impose various constraints to restrict on the basis set, and this
offers an advantage in optimizing the size of the Hamiltonian matrix.
The generation of the Hamiltonian matrix is also optimized by identifying
the pairs of states which have nonzero matrix elements. For the case of
the calculations with constraints, we use the bisection method to accelerate 
the identification and calculation of the nonzero Hamiltonian matrix elements.
All of the constituent steps are parallelizable. In our
implementation we have parallelized both the basis set generation and
calculation of the Hamiltonian matrix elements. Further more, we 
parallelize the diagonalization of the Hamiltonian matrix using 
{\sc PARPACK} \cite{maschhoff_96}. We benchmark our ED method 
with calculations of some ground-state properties of BHM.
Using the method, we show how to 
partition a lattice system, compute the reduced density matrix and 
calculate the topological entanglement entropy. This is applied to the 
$\nu=1/2$ FQH state.

\section{Acknowledgements}

The results presented in this paper were computed on Vikram-100, the 100TFLOP
HPC cluster and Param Vikram-1000 HPC cluster at Physical Research Laboratory,
Ahmedabad, India. We thank Dr. Rukmani Bai and Dr. Soumik Bandyopadhyay for  
useful discussions. D.A. would like to acknowledge support from the Science and 
Engineering Research Board, Department of Science and Technology, Government 
of India, through Project No. CRG/2022/007099 and support from the UGC through
the SAP (DRS-II), Department of Physics, Manipur University. 

\appendix


\section{Example illustration of row states, multirow states and basis-states
construction}
\label{appendA}

As an example, consider the states corresponding to an extra particle over the 
commensurate unit filling state. For this case, we consider a system of $10$ 
bosons on a $3 \times 3$ lattice. In order to generate the basis states for 
this system, we first construct the row states corresponding to the rows 
comprising of $3$ lattice sites. Let us constraint the single-site occupancies 
such that it can be either 1 or 2. In addition, we filter the possible 
row-state configurations by imposing the constraint on the total number of 
particles within a row as either 3 or 4. The possible row-state configurations 
with these constraints, $\eta = 1$, $N_B =3$, and $\sigma = 3$ and 
$\sigma + \delta =4$ are listed in Table~\ref{table_rows}, where, the 
row-state configurations $\ket{\phi_m} \equiv \ket {n_1,n_2,n_3}$, 
total $\beta = 4$ in number, can be uniquely identified by the row-state 
quantum number $m$. It is evident that the quantum number $m$ does not vary 
uniformly, and is thus unsuitable as labels of the row states in further 
computations. So, a uniformly spaced index $i$ is used to tag
the row states. As shown in Table~\ref{table_rows}, the index $i$ is ordered
as per the sequential ordering in quantum number $m$ such that $i = 1$
represents row state with the lowest value of $m$, $i = 2$ corresponds to the
row state with the second lowest value of $m$, and so on.

\begin{table}[ht]
 \begin{tabular}{ |c|c|c|c|c| }
  \hline
	 $\Ket{\phi_m} \equiv$      &   &  & &          \\
   $\big\vert\begin{matrix} n_1& n_2& n_3 \end{matrix}\big\rangle $ &
   $\big\vert\begin{matrix}   1&   1&   1 \end{matrix}\big\rangle $ &
   $\big\vert\begin{matrix}   2&   1&   1 \end{matrix}\big\rangle $ &
   $\big\vert\begin{matrix}   1&   2&   1 \end{matrix}\big\rangle $ &
   $\big\vert\begin{matrix}   1&   1&   2 \end{matrix}\big\rangle $ \\
  \hline                               
	 $m$         & $13$ & $14$ & $16$ & $22$         \\
  \hline                               
	 $i$         & $1$  & $2$  & $3$  & $\beta = 4$\\
  \hline                               
 \end{tabular}
 \caption{Table showing all possible row-state configurations together with
          the corresponding values of the row-state quantum number $m$ and
          state index $i$.}
 \label{table_rows}
\end{table}

Next, we consider the two-row states $\Ket{\Phi^2_{M}}$ generated from
the direct product of the row states $\ket{\phi_{m_1}}$ and $\ket{\phi_{m_2}}$,
chosen from the row-state configurations shown in Table~\ref{table_rows}.
At this stage, the possible two-row states can be filtered by constraining the
total number of bosons within the two rows to be atmost 10, since this is the
maximum number of bosons within the $3 \times 3$ lattice. With this constraint,
all possible two-row states configurations are $\beta^{(2)} =16$ in number,
and are shown in Table~\ref{table_clus}. Similar to the case of previously
constructed row states, the quantum number $M$ does not vary uniformly and is
thus not used for enumeration purpose. Instead, we tag each of the two-row
states with a uniformly spaced index $I$, which is sequenced in accordance
with the quantum numbers of the two contributing row states $(m_1,m_2)$, with
$m_2$ as the faster varying index.

\begin{table}[ht]
 \begin{tabular}{ |c|c|c|c| }
 \hline
 $\Ket{\Phi^2_{M}} = \begin{matrix} \Ket{\phi_{m_2}} \\ \otimes \\
 \ket{\phi_{m_1}} \end{matrix}$  & ${\bf M} = (m_1,m_2)$ 
 & $M = \Vert {\bf M} \Vert$ & $I$ \\  
         \hline
 $\bigg\vert\left.\begin{matrix} 1 & 1 & 1 \\ 1 & 1 & 1 \end{matrix}\right.
 \bigg\rangle$    &  $(13,13)$    &   364    &  $1$                \\  
	 \hline
 $\bigg\vert\left.\begin{matrix} 2 & 1 & 1 \\ 1 & 1 & 1 \end{matrix}\right.
 \bigg\rangle$    &  $(13,14)$    &   365    &  $2$                \\  
	 \hline
 $\bigg\vert\left.\begin{matrix} 1 & 2 & 1 \\ 1 & 1 & 1 \end{matrix}\right.
 \bigg\rangle$    &  $(13,16)$    &   367    &  $3$                \\  
	 \hline
 $\bigg\vert\left.\begin{matrix} 1 & 1 & 2 \\ 1 & 1 & 1 \end{matrix}\right.
 \bigg\rangle$    &  $(13,22)$    &   373    &  $4$                \\  
	 \hline
 $\bigg\vert\left.\begin{matrix} 1 & 1 & 1 \\ 2 & 1 & 1 \end{matrix}\right.
 \bigg\rangle$    &  $(14,13)$    &   391    &  $5$                \\  
	 \hline
 $\bigg\vert\left.\begin{matrix} 2 & 1 & 1 \\ 2 & 1 & 1 \end{matrix}\right.
 \bigg\rangle$    &  $(14,14)$    &   392    &  $6$                \\  
	 \hline
 $\bigg\vert\left.\begin{matrix} 1 & 2 & 1 \\ 2 & 1 & 1 \end{matrix}\right.
 \bigg\rangle$    &  $(14,16)$    &   394    &  $7$                \\  
	 \hline 
 $\bigg\vert\left.\begin{matrix} 1 & 1 & 2 \\ 2 & 1 & 1 \end{matrix}\right.
 \bigg\rangle$    &  $(14,22)$    &   400    &  $8$                \\  
	 \hline
 $\bigg\vert\left.\begin{matrix} 1 & 1 & 1 \\ 1 & 2 & 1 \end{matrix}\right.
 \bigg\rangle$    &  $(16,13)$    &   445    &  $9$                \\  
	 \hline
 $\bigg\vert\left.\begin{matrix} 2 & 1 & 1 \\ 1 & 2 & 1 \end{matrix}\right.
 \bigg\rangle$    &  $(16,14)$    &   446    &  $10$               \\  
	 \hline
 $\bigg\vert\left.\begin{matrix} 1 & 2 & 1 \\ 1 & 2 & 1 \end{matrix}\right.
 \bigg\rangle$    &  $(16,16)$    &   448    &  $11$               \\  
	 \hline
 $\bigg\vert\left.\begin{matrix} 1 & 1 & 2 \\ 1 & 2 & 1 \end{matrix}\right.
 \bigg\rangle$    &  $(16,22)$    &   454    &  $12$               \\  
	 \hline
 $\bigg\vert\left.\begin{matrix} 1 & 1 & 1 \\ 1 & 1 & 2 \end{matrix}\right.
 \bigg\rangle$    &  $(22,13)$    &   607    &  $13$               \\  
	 \hline
 $\bigg\vert\left.\begin{matrix} 2 & 1 & 1 \\ 1 & 1 & 2 \end{matrix}\right.
 \bigg\rangle$    &  $(22,14)$    &   608    &  $14$               \\  
	 \hline
 $\bigg\vert\left.\begin{matrix} 1 & 2 & 1 \\ 1 & 1 & 2 \end{matrix}\right.
 \bigg\rangle$    &  $(22,16)$    &   610    &  $15$               \\  
	 \hline
 $\bigg\vert\left.\begin{matrix} 1 & 1 & 2 \\ 1 & 1 & 2 \end{matrix}\right.
 \bigg\rangle$    &  $(22,22)$    &   616    &  $\beta^{(2)} =16$ \\  
	 \hline
 \end{tabular}
 \caption{Table showing all possible two-row states configurations with the
          corresponding vector label ${\bf M}$ for the constraint of atmost
          10 particles within 2 rows. The quantum number 
	  $M = \Vert{\bf M} \Vert$ is given by Eq.~(\ref{int_label}).}
 \label{table_clus}
\end{table}

Finally, the basis states corresponding to the system of $N=10$ particles on a
$3 \times 3$ lattice is constructed from the direct product of the rowstate
$\ket{\phi_{m_3}}$ and the two-row state $\Ket{\Phi^2_{M}}$, constructed
earlier and shown in Table~\ref{table_rows} and \ref{table_clus}. Out of all
possible combinations, valid basis states are enumerated by filtering out only
those states for which the total number of particles in the $3 \times 3$
lattice is fixed as $N = 10$. Table~\ref{table_basis} lists all
such possible basis states, which are $\Gamma$ in number. Again, similar to the
previous cases, the basis states are tagged with uniformly varying index $I$
which is sequenced in accordance with the quantum numbers of contributing
row states $(m_1, m_2, m_3)$, with $m_1$ being slowest varying while $m_3$ as
fastest varying index. 

\begin{table}[ht]
 \begin{tabular}{ |c|c|c|c| }
  \hline
 $\Ket{\Phi^3_{M}} = \begin{matrix} \Ket{\phi_{m_3}} \\ \otimes
 \\ \ket{\phi_{m_2}} \\ \otimes \\ \ket{\phi_{m_1}} \end{matrix}$
 & ${\bf M} = (m_1,m_2, m_3)$ & $M = \Vert {\bf M} \Vert$ & $\quad I \quad$\\
	 \hline 
 $\Bigg\vert\left.\begin{matrix} 2 & 1 & 1 \\ 1 & 1 & 1 \\ 1 & 1 & 1
 \end{matrix}\right.\Bigg\rangle$   &  $(13,13,14)$   &  9842   &  $1$   \\
	 \hline 
 $\Bigg\vert\left.\begin{matrix} 1 & 2 & 1 \\ 1 & 1 & 1 \\ 1 & 1 & 1
 \end{matrix}\right.\Bigg\rangle$   &  $(13,13,16)$   &  9844   &  $2$   \\
	 \hline 
 $\Bigg\vert\left.\begin{matrix} 1 & 1 & 2 \\ 1 & 1 & 1 \\ 1 & 1 & 1
 \end{matrix}\right.\Bigg\rangle$   &  $(13,13,22)$   &  9850   &  $3$   \\
	 \hline 
 $\Bigg\vert\left.\begin{matrix} 1 & 1 & 1 \\ 2 & 1 & 1 \\ 1 & 1 & 1
 \end{matrix}\right.\Bigg\rangle$   &  $(13,14,13)$   &  9868   &  $4$   \\
	 \hline 
 $\Bigg\vert\left.\begin{matrix} 1 & 1 & 1 \\ 1 & 2 & 1 \\ 1 & 1 & 1
 \end{matrix}\right.\Bigg\rangle$   &  $(13,16,13)$   &  9922   &  $5$   \\
	 \hline 
 $\Bigg\vert\left.\begin{matrix} 1 & 1 & 1 \\ 1 & 1 & 2 \\ 1 & 1 & 1
 \end{matrix}\right.\Bigg\rangle$   &  $(13,22,13)$   & 10084   &  $6$   \\
	 \hline 
 $\Bigg\vert\left.\begin{matrix} 1 & 1 & 1 \\ 1 & 1 & 1 \\ 2 & 1 & 1
 \end{matrix}\right.\Bigg\rangle$   &  $(14,13,13)$   & 10570   &  $7$   \\
	 \hline 
 $\Bigg\vert\left.\begin{matrix} 1 & 1 & 1 \\ 1 & 1 & 1 \\ 1 & 2 & 1
 \end{matrix}\right.\Bigg\rangle$   &  $(16,13,13)$   & 12028   &  $8$   \\
	 \hline 
 $\Bigg\vert\left.\begin{matrix} 1 & 1 & 1 \\ 1 & 1 & 1 \\ 1 & 1 & 2
 \end{matrix}\right.\Bigg\rangle$   &  $(22,13,13)$   & 16402   & $\Gamma =9$\\
	 \hline                               
 \end{tabular}
 \caption{Table showing all possible basis states identified by the 
          corresponding vector label ${\bf M}$ for a system of ten particles
          on a $3 \times 3$ lattice.}
 \label{table_basis}
\end{table}

\section{Additional details regarding the $T_{\rm ED}$ computations}
\label{appendB}

For comparison of $T_{\rm ED}$ shown in Fig.~\ref{comparison_ED} and 
\ref{time_stred}, the codes were executed on a cluster computer with 
256GB RAM per node. In the Fig.~\ref{comparison_ED}, we compare the execution
time of our ED code (written in {\sc FORTRAN}) against the ``quantum basis" 
code. The later code is written in C language and utilizes the concept of 
``Lin Tables" for basis identification. And, the matrix diagonalization is
possible with either Lanczos method or with implicitly restarted Arnoldi 
method using {\sc ARPACK} package. Using the Lanczos diagonalization procedure 
in ``quantum basis", we note that the ground state is identical to the value 
obtained from our method. For comparing the two methods, we consider the
$T_{\rm ED}$ using the {\sc ARPACK} method. For the results shown in 
Fig.~\ref{comparison_ED}, the two codes were executed sequentially. And, 
for $T_{\rm ED}$ comparison shown in Fig.~\ref{time_stred}, the ED code was 
executed on two threads.



\bibliography{ref}{}

\begin{thebibliography}{48}%
\makeatletter
\providecommand \@ifxundefined [1]{%
 \@ifx{#1\undefined}
}%
\providecommand \@ifnum [1]{%
 \ifnum #1\expandafter \@firstoftwo
 \else \expandafter \@secondoftwo
 \fi
}%
\providecommand \@ifx [1]{%
 \ifx #1\expandafter \@firstoftwo
 \else \expandafter \@secondoftwo
 \fi
}%
\providecommand \natexlab [1]{#1}%
\providecommand \enquote  [1]{``#1''}%
\providecommand \bibnamefont  [1]{#1}%
\providecommand \bibfnamefont [1]{#1}%
\providecommand \citenamefont [1]{#1}%
\providecommand \href@noop [0]{\@secondoftwo}%
\providecommand \href [0]{\begingroup \@sanitize@url \@href}%
\providecommand \@href[1]{\@@startlink{#1}\@@href}%
\providecommand \@@href[1]{\endgroup#1\@@endlink}%
\providecommand \@sanitize@url [0]{\catcode `\\12\catcode `\$12\catcode
  `\&12\catcode `\#12\catcode `\^12\catcode `\_12\catcode `\%12\relax}%
\providecommand \@@startlink[1]{}%
\providecommand \@@endlink[0]{}%
\providecommand \url  [0]{\begingroup\@sanitize@url \@url }%
\providecommand \@url [1]{\endgroup\@href {#1}{\urlprefix }}%
\providecommand \urlprefix  [0]{URL }%
\providecommand \Eprint [0]{\href }%
\providecommand \doibase [0]{http://dx.doi.org/}%
\providecommand \selectlanguage [0]{\@gobble}%
\providecommand \bibinfo  [0]{\@secondoftwo}%
\providecommand \bibfield  [0]{\@secondoftwo}%
\providecommand \translation [1]{[#1]}%
\providecommand \BibitemOpen [0]{}%
\providecommand \bibitemStop [0]{}%
\providecommand \bibitemNoStop [0]{.\EOS\space}%
\providecommand \EOS [0]{\spacefactor3000\relax}%
\providecommand \BibitemShut  [1]{\csname bibitem#1\endcsname}%
\let\auto@bib@innerbib\@empty
\bibitem [{\citenamefont {Chaikin}\ and\ \citenamefont
  {Lubensky}(1995)}]{chaikin_95}%
  \BibitemOpen
  \bibfield  {author} {\bibinfo {author} {\bibfnamefont {P.~M.}\ \bibnamefont
  {Chaikin}}\ and\ \bibinfo {author} {\bibfnamefont {T.~C.}\ \bibnamefont
  {Lubensky}},\ }\href {\doibase 10.1017/CBO9780511813467} {\emph {\bibinfo
  {title} {Principles of Condensed Matter Physics}}}\ (\bibinfo  {publisher}
  {Cambridge University Press},\ \bibinfo {year} {1995})\BibitemShut {NoStop}%
\bibitem [{\citenamefont {Sandvik}\ and\ \citenamefont
  {Kurkij\"arvi}(1991)}]{sandvik_91}%
  \BibitemOpen
  \bibfield  {author} {\bibinfo {author} {\bibfnamefont {Anders~W.}\
  \bibnamefont {Sandvik}}\ and\ \bibinfo {author} {\bibfnamefont {Juhani}\
  \bibnamefont {Kurkij\"arvi}},\ }\bibfield  {title} {\enquote {\bibinfo
  {title} {Quantum monte carlo simulation method for spin systems},}\ }\href
  {\doibase 10.1103/PhysRevB.43.5950} {\bibfield  {journal} {\bibinfo
  {journal} {Phys. Rev. B}\ }\textbf {\bibinfo {volume} {43}},\ \bibinfo
  {pages} {5950--5961} (\bibinfo {year} {1991})}\BibitemShut {NoStop}%
\bibitem [{\citenamefont {Sandvik}(1992)}]{sandvik_92}%
  \BibitemOpen
  \bibfield  {author} {\bibinfo {author} {\bibfnamefont {A~W}\ \bibnamefont
  {Sandvik}},\ }\bibfield  {title} {\enquote {\bibinfo {title} {A
  generalization of handscomb's quantum monte carlo scheme-application to the
  1d hubbard model},}\ }\href {\doibase 10.1088/0305-4470/25/13/017} {\bibfield
   {journal} {\bibinfo  {journal} {Journal of Physics A: Mathematical and
  General}\ }\textbf {\bibinfo {volume} {25}},\ \bibinfo {pages} {3667}
  (\bibinfo {year} {1992})}\BibitemShut {NoStop}%
\bibitem [{\citenamefont {Henelius}\ and\ \citenamefont
  {Sandvik}(2000)}]{henelius_00}%
  \BibitemOpen
  \bibfield  {author} {\bibinfo {author} {\bibfnamefont {Patrik}\ \bibnamefont
  {Henelius}}\ and\ \bibinfo {author} {\bibfnamefont {Anders~W.}\ \bibnamefont
  {Sandvik}},\ }\bibfield  {title} {\enquote {\bibinfo {title} {Sign problem in
  monte carlo simulations of frustrated quantum spin systems},}\ }\href
  {\doibase 10.1103/PhysRevB.62.1102} {\bibfield  {journal} {\bibinfo
  {journal} {Phys. Rev. B}\ }\textbf {\bibinfo {volume} {62}},\ \bibinfo
  {pages} {1102--1113} (\bibinfo {year} {2000})}\BibitemShut {NoStop}%
\bibitem [{\citenamefont {Lanczos}(1952)}]{lanczos_1952}%
  \BibitemOpen
  \bibfield  {author} {\bibinfo {author} {\bibfnamefont {Cornelius}\
  \bibnamefont {Lanczos}},\ }\bibfield  {title} {\enquote {\bibinfo {title}
  {Solution of systems of linear equations by minimized iterations},}\
  }\href@noop {} {\bibfield  {journal} {\bibinfo  {journal} {J. Res. Nat. Bur.
  Standards}\ }\textbf {\bibinfo {volume} {49}},\ \bibinfo {pages} {33--53}
  (\bibinfo {year} {1952})}\BibitemShut {NoStop}%
\bibitem [{\citenamefont {Calvetti}\ \emph {et~al.}(1994)\citenamefont
  {Calvetti}, \citenamefont {Reichel},\ and\ \citenamefont
  {Sorensen}}]{calvetti_94}%
  \BibitemOpen
  \bibfield  {author} {\bibinfo {author} {\bibfnamefont {D.}~\bibnamefont
  {Calvetti}}, \bibinfo {author} {\bibfnamefont {L.}~\bibnamefont {Reichel}}, \
  and\ \bibinfo {author} {\bibfnamefont {D.~C.}\ \bibnamefont {Sorensen}},\
  }\bibfield  {title} {\enquote {\bibinfo {title} {An implicitly restarted
  lanczos method for large symmetric eigenvalue problems},}\ }\href@noop {}
  {\bibfield  {journal} {\bibinfo  {journal} {Electron. Trans. Numer. Anal.}\
  }\textbf {\bibinfo {volume} {2}},\ \bibinfo {pages} {1--21} (\bibinfo {year}
  {1994})}\BibitemShut {NoStop}%
\bibitem [{\citenamefont {L\"auchli}\ \emph {et~al.}(2011)\citenamefont
  {L\"auchli}, \citenamefont {Sudan},\ and\ \citenamefont
  {S\o{}rensen}}]{lauchli_11}%
  \BibitemOpen
  \bibfield  {author} {\bibinfo {author} {\bibfnamefont {Andreas~M.}\
  \bibnamefont {L\"auchli}}, \bibinfo {author} {\bibfnamefont {Julien}\
  \bibnamefont {Sudan}}, \ and\ \bibinfo {author} {\bibfnamefont {Erik~S.}\
  \bibnamefont {S\o{}rensen}},\ }\bibfield  {title} {\enquote {\bibinfo {title}
  {Ground-state energy and spin gap of spin-$\frac{1}{2}$ kagom\'e-heisenberg
  antiferromagnetic clusters: Large-scale exact diagonalization results},}\
  }\href {\doibase 10.1103/PhysRevB.83.212401} {\bibfield  {journal} {\bibinfo
  {journal} {Phys. Rev. B}\ }\textbf {\bibinfo {volume} {83}},\ \bibinfo
  {pages} {212401} (\bibinfo {year} {2011})}\BibitemShut {NoStop}%
\bibitem [{\citenamefont {L\"auchli}\ \emph {et~al.}(2019)\citenamefont
  {L\"auchli}, \citenamefont {Sudan},\ and\ \citenamefont
  {Moessner}}]{lauchli_19}%
  \BibitemOpen
  \bibfield  {author} {\bibinfo {author} {\bibfnamefont {Andreas~M.}\
  \bibnamefont {L\"auchli}}, \bibinfo {author} {\bibfnamefont {Julien}\
  \bibnamefont {Sudan}}, \ and\ \bibinfo {author} {\bibfnamefont {Roderich}\
  \bibnamefont {Moessner}},\ }\bibfield  {title} {\enquote {\bibinfo {title}
  {$s=\frac{1}{2}$ kagome heisenberg antiferromagnet revisited},}\ }\href
  {\doibase 10.1103/PhysRevB.100.155142} {\bibfield  {journal} {\bibinfo
  {journal} {Phys. Rev. B}\ }\textbf {\bibinfo {volume} {100}},\ \bibinfo
  {pages} {155142} (\bibinfo {year} {2019})}\BibitemShut {NoStop}%
\bibitem [{\citenamefont {Bai}\ \emph {et~al.}(2018)\citenamefont {Bai},
  \citenamefont {Bandyopadhyay}, \citenamefont {Pal}, \citenamefont {Suthar},\
  and\ \citenamefont {Angom}}]{bai_18}%
  \BibitemOpen
  \bibfield  {author} {\bibinfo {author} {\bibfnamefont {Rukmani}\ \bibnamefont
  {Bai}}, \bibinfo {author} {\bibfnamefont {Soumik}\ \bibnamefont
  {Bandyopadhyay}}, \bibinfo {author} {\bibfnamefont {Sukla}\ \bibnamefont
  {Pal}}, \bibinfo {author} {\bibfnamefont {K.}~\bibnamefont {Suthar}}, \ and\
  \bibinfo {author} {\bibfnamefont {D.}~\bibnamefont {Angom}},\ }\bibfield
  {title} {\enquote {\bibinfo {title} {Bosonic quantum hall states in
  single-layer two-dimensional optical lattices},}\ }\href {\doibase
  10.1103/PhysRevA.98.023606} {\bibfield  {journal} {\bibinfo  {journal} {Phys.
  Rev. A}\ }\textbf {\bibinfo {volume} {98}},\ \bibinfo {pages} {023606}
  (\bibinfo {year} {2018})}\BibitemShut {NoStop}%
\bibitem [{\citenamefont {Lewenstein}\ \emph {et~al.}(2012)\citenamefont
  {Lewenstein}, \citenamefont {Sanpera},\ and\ \citenamefont
  {Ahufinger}}]{lewenstein_12}%
  \BibitemOpen
  \bibfield  {author} {\bibinfo {author} {\bibfnamefont {Maciej}\ \bibnamefont
  {Lewenstein}}, \bibinfo {author} {\bibfnamefont {Anna}\ \bibnamefont
  {Sanpera}}, \ and\ \bibinfo {author} {\bibfnamefont {Verònica}\ \bibnamefont
  {Ahufinger}},\ }\href {\doibase 10.1093/acprof:oso/9780199573127.001.0001}
  {\emph {\bibinfo {title} {{Ultracold Atoms in Optical Lattices: Simulating
  quantum many-body systems}}}}\ (\bibinfo  {publisher} {Oxford University
  Press},\ \bibinfo {year} {2012})\BibitemShut {NoStop}%
\bibitem [{\citenamefont {Greiner}\ \emph {et~al.}(2002)\citenamefont
  {Greiner}, \citenamefont {Mandel}, \citenamefont {Esslinger}, \citenamefont
  {H\"ansch},\ and\ \citenamefont {Bloch}}]{greiner_02}%
  \BibitemOpen
  \bibfield  {author} {\bibinfo {author} {\bibfnamefont {M.}~\bibnamefont
  {Greiner}}, \bibinfo {author} {\bibfnamefont {O.}~\bibnamefont {Mandel}},
  \bibinfo {author} {\bibfnamefont {T.}~\bibnamefont {Esslinger}}, \bibinfo
  {author} {\bibfnamefont {T.~W.}\ \bibnamefont {H\"ansch}}, \ and\ \bibinfo
  {author} {\bibfnamefont {I.}~\bibnamefont {Bloch}},\ }\bibfield  {title}
  {\enquote {\bibinfo {title} {Quantum phase transition from a superfluid to a
  {M}ott insulator in a gas of ultracold atoms},}\ }\href {\doibase
  10.1038/415039a} {\bibfield  {journal} {\bibinfo  {journal} {Nature
  (London)}\ }\textbf {\bibinfo {volume} {415}},\ \bibinfo {pages} {39}
  (\bibinfo {year} {2002})}\BibitemShut {NoStop}%
\bibitem [{\citenamefont {Lewenstein}\ \emph {et~al.}(2007)\citenamefont
  {Lewenstein}, \citenamefont {Sanpera}, \citenamefont {Ahufinger},
  \citenamefont {Damski}, \citenamefont {Sen(De)},\ and\ \citenamefont
  {Sen}}]{lewenstein_07}%
  \BibitemOpen
  \bibfield  {author} {\bibinfo {author} {\bibfnamefont {M.}~\bibnamefont
  {Lewenstein}}, \bibinfo {author} {\bibfnamefont {A.}~\bibnamefont {Sanpera}},
  \bibinfo {author} {\bibfnamefont {V.}~\bibnamefont {Ahufinger}}, \bibinfo
  {author} {\bibfnamefont {B.}~\bibnamefont {Damski}}, \bibinfo {author}
  {\bibfnamefont {A.}~\bibnamefont {Sen(De)}}, \ and\ \bibinfo {author}
  {\bibfnamefont {U.}~\bibnamefont {Sen}},\ }\bibfield  {title} {\enquote
  {\bibinfo {title} {Ultracold atomic gases in optical lattices: mimicking
  condensed matter physics and beyond},}\ }\href {\doibase
  10.1080/00018730701223200} {\bibfield  {journal} {\bibinfo  {journal} {Adv.
  Phys.}\ }\textbf {\bibinfo {volume} {56}},\ \bibinfo {pages} {243} (\bibinfo
  {year} {2007})}\BibitemShut {NoStop}%
\bibitem [{\citenamefont {Hubbard}(1963)}]{hubbard_63}%
  \BibitemOpen
  \bibfield  {author} {\bibinfo {author} {\bibfnamefont {J.}~\bibnamefont
  {Hubbard}},\ }\bibfield  {title} {\enquote {\bibinfo {title} {Electron
  correlations in narrow energy bands},}\ }\href {\doibase
  10.1098/rspa.1963.0204} {\bibfield  {journal} {\bibinfo  {journal} {Proc.
  Royal Soc. A}\ }\textbf {\bibinfo {volume} {276}},\ \bibinfo {pages} {238}
  (\bibinfo {year} {1963})}\BibitemShut {NoStop}%
\bibitem [{\citenamefont {Fisher}\ \emph {et~al.}(1989)\citenamefont {Fisher},
  \citenamefont {Weichman}, \citenamefont {Grinstein},\ and\ \citenamefont
  {Fisher}}]{fisher_89}%
  \BibitemOpen
  \bibfield  {author} {\bibinfo {author} {\bibfnamefont {M.~P.~A.}\
  \bibnamefont {Fisher}}, \bibinfo {author} {\bibfnamefont {P.~B.}\
  \bibnamefont {Weichman}}, \bibinfo {author} {\bibfnamefont {G.}~\bibnamefont
  {Grinstein}}, \ and\ \bibinfo {author} {\bibfnamefont {D.~S.}\ \bibnamefont
  {Fisher}},\ }\bibfield  {title} {\enquote {\bibinfo {title} {Boson
  localization and the superfluid-insulator transition},}\ }\href {\doibase
  10.1103/PhysRevB.40.546} {\bibfield  {journal} {\bibinfo  {journal} {Phys.
  Rev. B}\ }\textbf {\bibinfo {volume} {40}},\ \bibinfo {pages} {546} (\bibinfo
  {year} {1989})}\BibitemShut {NoStop}%
\bibitem [{\citenamefont {Jaksch}\ \emph {et~al.}(1998)\citenamefont {Jaksch},
  \citenamefont {Bruder}, \citenamefont {Cirac}, \citenamefont {Gardiner},\
  and\ \citenamefont {Zoller}}]{jaksch_98}%
  \BibitemOpen
  \bibfield  {author} {\bibinfo {author} {\bibfnamefont {D.}~\bibnamefont
  {Jaksch}}, \bibinfo {author} {\bibfnamefont {C.}~\bibnamefont {Bruder}},
  \bibinfo {author} {\bibfnamefont {J.~I.}\ \bibnamefont {Cirac}}, \bibinfo
  {author} {\bibfnamefont {C.~W.}\ \bibnamefont {Gardiner}}, \ and\ \bibinfo
  {author} {\bibfnamefont {P.}~\bibnamefont {Zoller}},\ }\bibfield  {title}
  {\enquote {\bibinfo {title} {Cold bosonic atoms in optical lattices},}\
  }\href {\doibase 10.1103/PhysRevLett.81.3108} {\bibfield  {journal} {\bibinfo
   {journal} {Phys. Rev. Lett.}\ }\textbf {\bibinfo {volume} {81}},\ \bibinfo
  {pages} {3108} (\bibinfo {year} {1998})}\BibitemShut {NoStop}%
\bibitem [{\citenamefont {Rokhsar}\ and\ \citenamefont
  {Kotliar}(1991)}]{rokhsar_91}%
  \BibitemOpen
  \bibfield  {author} {\bibinfo {author} {\bibfnamefont {D.~S.}\ \bibnamefont
  {Rokhsar}}\ and\ \bibinfo {author} {\bibfnamefont {B.~G.}\ \bibnamefont
  {Kotliar}},\ }\bibfield  {title} {\enquote {\bibinfo {title} {Gutzwiller
  projection for bosons},}\ }\href {\doibase 10.1103/PhysRevB.44.10328}
  {\bibfield  {journal} {\bibinfo  {journal} {Phys. Rev. B}\ }\textbf {\bibinfo
  {volume} {44}},\ \bibinfo {pages} {10328} (\bibinfo {year}
  {1991})}\BibitemShut {NoStop}%
\bibitem [{\citenamefont {Sheshadri}\ \emph {et~al.}(1993)\citenamefont
  {Sheshadri}, \citenamefont {Krishnamurthy}, \citenamefont {Pandit},\ and\
  \citenamefont {Ramakrishnan}}]{sheshadri_93}%
  \BibitemOpen
  \bibfield  {author} {\bibinfo {author} {\bibfnamefont {K.}~\bibnamefont
  {Sheshadri}}, \bibinfo {author} {\bibfnamefont {H.~R.}\ \bibnamefont
  {Krishnamurthy}}, \bibinfo {author} {\bibfnamefont {R.}~\bibnamefont
  {Pandit}}, \ and\ \bibinfo {author} {\bibfnamefont {T.~V.}\ \bibnamefont
  {Ramakrishnan}},\ }\bibfield  {title} {\enquote {\bibinfo {title} {Superfluid
  and insulating phases in an interacting-boson model: Mean-field theory and
  the {RPA}},}\ }\href {\doibase https://doi.org/10.1209/0295-5075/22/4/004}
  {\bibfield  {journal} {\bibinfo  {journal} {EPL}\ }\textbf {\bibinfo {volume}
  {22}},\ \bibinfo {pages} {257} (\bibinfo {year} {1993})}\BibitemShut
  {NoStop}%
\bibitem [{\citenamefont {Oktel}\ \emph {et~al.}(2007)\citenamefont {Oktel},
  \citenamefont {Ni\ifmmode \mbox{\c{t}}\else \c{t}\fi{} \ifmmode~\u{a}\else
  \u{a}\fi{}},\ and\ \citenamefont {Tanatar}}]{oktel_07}%
  \BibitemOpen
  \bibfield  {author} {\bibinfo {author} {\bibfnamefont {M.~\"O.}\ \bibnamefont
  {Oktel}}, \bibinfo {author} {\bibfnamefont {M.}~\bibnamefont {Ni\ifmmode
  \mbox{\c{t}}\else \c{t}\fi{} \ifmmode~\u{a}\else \u{a}\fi{}}}, \ and\
  \bibinfo {author} {\bibfnamefont {B.}~\bibnamefont {Tanatar}},\ }\bibfield
  {title} {\enquote {\bibinfo {title} {Mean-field theory for {B}ose-{H}ubbard
  model under a magnetic field},}\ }\href {\doibase 10.1103/PhysRevB.75.045133}
  {\bibfield  {journal} {\bibinfo  {journal} {Phys. Rev. B}\ }\textbf {\bibinfo
  {volume} {75}},\ \bibinfo {pages} {045133} (\bibinfo {year}
  {2007})}\BibitemShut {NoStop}%
\bibitem [{\citenamefont {Aidelsburger}\ \emph {et~al.}(2011)\citenamefont
  {Aidelsburger}, \citenamefont {Atala}, \citenamefont {Nascimb\`ene},
  \citenamefont {Trotzky}, \citenamefont {Chen},\ and\ \citenamefont
  {Bloch}}]{aidelsburger_11}%
  \BibitemOpen
  \bibfield  {author} {\bibinfo {author} {\bibfnamefont {M.}~\bibnamefont
  {Aidelsburger}}, \bibinfo {author} {\bibfnamefont {M.}~\bibnamefont {Atala}},
  \bibinfo {author} {\bibfnamefont {S.}~\bibnamefont {Nascimb\`ene}}, \bibinfo
  {author} {\bibfnamefont {S.}~\bibnamefont {Trotzky}}, \bibinfo {author}
  {\bibfnamefont {Y.-A.}\ \bibnamefont {Chen}}, \ and\ \bibinfo {author}
  {\bibfnamefont {I.}~\bibnamefont {Bloch}},\ }\bibfield  {title} {\enquote
  {\bibinfo {title} {Experimental realization of strong effective magnetic
  fields in an optical lattice},}\ }\href {\doibase
  10.1103/PhysRevLett.107.255301} {\bibfield  {journal} {\bibinfo  {journal}
  {Phys. Rev. Lett.}\ }\textbf {\bibinfo {volume} {107}},\ \bibinfo {pages}
  {255301} (\bibinfo {year} {2011})}\BibitemShut {NoStop}%
\bibitem [{\citenamefont {Aidelsburger}\ \emph {et~al.}(2013)\citenamefont
  {Aidelsburger}, \citenamefont {Atala}, \citenamefont {Lohse}, \citenamefont
  {Barreiro}, \citenamefont {Paredes},\ and\ \citenamefont
  {Bloch}}]{aidelsburger_13}%
  \BibitemOpen
  \bibfield  {author} {\bibinfo {author} {\bibfnamefont {M.}~\bibnamefont
  {Aidelsburger}}, \bibinfo {author} {\bibfnamefont {M.}~\bibnamefont {Atala}},
  \bibinfo {author} {\bibfnamefont {M.}~\bibnamefont {Lohse}}, \bibinfo
  {author} {\bibfnamefont {J.~T.}\ \bibnamefont {Barreiro}}, \bibinfo {author}
  {\bibfnamefont {B.}~\bibnamefont {Paredes}}, \ and\ \bibinfo {author}
  {\bibfnamefont {I.}~\bibnamefont {Bloch}},\ }\bibfield  {title} {\enquote
  {\bibinfo {title} {Realization of the {H}ofstadter {H}amiltonian with
  ultracold atoms in optical lattices},}\ }\href {\doibase
  10.1103/PhysRevLett.111.185301} {\bibfield  {journal} {\bibinfo  {journal}
  {Phys. Rev. Lett.}\ }\textbf {\bibinfo {volume} {111}},\ \bibinfo {pages}
  {185301} (\bibinfo {year} {2013})}\BibitemShut {NoStop}%
\bibitem [{\citenamefont {Jim\'enez-Garc\'{\i}a}\ \emph
  {et~al.}(2012)\citenamefont {Jim\'enez-Garc\'{\i}a}, \citenamefont {LeBlanc},
  \citenamefont {Williams}, \citenamefont {Beeler}, \citenamefont {Perry},\
  and\ \citenamefont {Spielman}}]{garcia_12}%
  \BibitemOpen
  \bibfield  {author} {\bibinfo {author} {\bibfnamefont {K.}~\bibnamefont
  {Jim\'enez-Garc\'{\i}a}}, \bibinfo {author} {\bibfnamefont {L.~J.}\
  \bibnamefont {LeBlanc}}, \bibinfo {author} {\bibfnamefont {R.~A.}\
  \bibnamefont {Williams}}, \bibinfo {author} {\bibfnamefont {M.~C.}\
  \bibnamefont {Beeler}}, \bibinfo {author} {\bibfnamefont {A.~R.}\
  \bibnamefont {Perry}}, \ and\ \bibinfo {author} {\bibfnamefont {I.~B.}\
  \bibnamefont {Spielman}},\ }\bibfield  {title} {\enquote {\bibinfo {title}
  {Peierls substitution in an engineered lattice potential},}\ }\href {\doibase
  10.1103/PhysRevLett.108.225303} {\bibfield  {journal} {\bibinfo  {journal}
  {Phys. Rev. Lett.}\ }\textbf {\bibinfo {volume} {108}},\ \bibinfo {pages}
  {225303} (\bibinfo {year} {2012})}\BibitemShut {NoStop}%
\bibitem [{\citenamefont {Dalibard}\ \emph {et~al.}(2011)\citenamefont
  {Dalibard}, \citenamefont {Gerbier}, \citenamefont
  {Juzeli\ifmmode~\bar{u}\else \={u}\fi{}nas},\ and\ \citenamefont
  {\"Ohberg}}]{dalibard_2011}%
  \BibitemOpen
  \bibfield  {author} {\bibinfo {author} {\bibfnamefont {Jean}\ \bibnamefont
  {Dalibard}}, \bibinfo {author} {\bibfnamefont {Fabrice}\ \bibnamefont
  {Gerbier}}, \bibinfo {author} {\bibfnamefont {Gediminas}\ \bibnamefont
  {Juzeli\ifmmode~\bar{u}\else \={u}\fi{}nas}}, \ and\ \bibinfo {author}
  {\bibfnamefont {Patrik}\ \bibnamefont {\"Ohberg}},\ }\bibfield  {title}
  {\enquote {\bibinfo {title} {Colloquium: Artificial gauge potentials for
  neutral atoms},}\ }\href {\doibase 10.1103/RevModPhys.83.1523} {\bibfield
  {journal} {\bibinfo  {journal} {Rev. Mod. Phys.}\ }\textbf {\bibinfo {volume}
  {83}},\ \bibinfo {pages} {1523--1543} (\bibinfo {year} {2011})}\BibitemShut
  {NoStop}%
\bibitem [{\citenamefont {Aidelsburger}(2018)}]{aidelsburger_2018}%
  \BibitemOpen
  \bibfield  {author} {\bibinfo {author} {\bibfnamefont {M}~\bibnamefont
  {Aidelsburger}},\ }\bibfield  {title} {\enquote {\bibinfo {title} {Artificial
  gauge fields and topology with ultracold atoms in optical lattices},}\ }\href
  {\doibase 10.1088/1361-6455/aac120} {\bibfield  {journal} {\bibinfo
  {journal} {Journal of Physics B: Atomic, Molecular and Optical Physics}\
  }\textbf {\bibinfo {volume} {51}},\ \bibinfo {pages} {193001} (\bibinfo
  {year} {2018})}\BibitemShut {NoStop}%
\bibitem [{\citenamefont {Cooper}\ \emph {et~al.}(2019)\citenamefont {Cooper},
  \citenamefont {Dalibard},\ and\ \citenamefont {Spielman}}]{cooper_2019}%
  \BibitemOpen
  \bibfield  {author} {\bibinfo {author} {\bibfnamefont {N.~R.}\ \bibnamefont
  {Cooper}}, \bibinfo {author} {\bibfnamefont {J.}~\bibnamefont {Dalibard}}, \
  and\ \bibinfo {author} {\bibfnamefont {I.~B.}\ \bibnamefont {Spielman}},\
  }\bibfield  {title} {\enquote {\bibinfo {title} {Topological bands for
  ultracold atoms},}\ }\href {\doibase 10.1103/RevModPhys.91.015005} {\bibfield
   {journal} {\bibinfo  {journal} {Rev. Mod. Phys.}\ }\textbf {\bibinfo
  {volume} {91}},\ \bibinfo {pages} {015005} (\bibinfo {year}
  {2019})}\BibitemShut {NoStop}%
\bibitem [{\citenamefont {Hauke}\ and\ \citenamefont
  {Carusotto}(2024)}]{hauke_2022}%
  \BibitemOpen
  \bibfield  {author} {\bibinfo {author} {\bibfnamefont {Philipp}\ \bibnamefont
  {Hauke}}\ and\ \bibinfo {author} {\bibfnamefont {Iacopo}\ \bibnamefont
  {Carusotto}},\ }\bibfield  {title} {\enquote {\bibinfo {title} {Quantum hall
  and synthetic magnetic-field effects in ultra-cold atomic systems},}\ }in\
  \href {\doibase https://doi.org/10.1016/B978-0-323-90800-9.00061-5} {\emph
  {\bibinfo {booktitle} {Encyclopedia of Condensed Matter Physics (Second
  Edition)}}},\ \bibinfo {editor} {edited by\ \bibinfo {editor} {\bibfnamefont
  {Tapash}\ \bibnamefont {Chakraborty}}}\ (\bibinfo  {publisher} {Academic
  Press},\ \bibinfo {address} {Oxford},\ \bibinfo {year} {2024})\ \bibinfo
  {edition} {second edition}\ ed.,\ pp.\ \bibinfo {pages}
  {629--639}\BibitemShut {NoStop}%
\bibitem [{\citenamefont {L{\'e}onard}\ \emph {et~al.}(2023)\citenamefont
  {L{\'e}onard}, \citenamefont {Kim}, \citenamefont {Kwan}, \citenamefont
  {Segura}, \citenamefont {Grusdt}, \citenamefont {Repellin}, \citenamefont
  {Goldman},\ and\ \citenamefont {Greiner}}]{leonard_23}%
  \BibitemOpen
  \bibfield  {author} {\bibinfo {author} {\bibfnamefont {Julian}\ \bibnamefont
  {L{\'e}onard}}, \bibinfo {author} {\bibfnamefont {Sooshin}\ \bibnamefont
  {Kim}}, \bibinfo {author} {\bibfnamefont {Joyce}\ \bibnamefont {Kwan}},
  \bibinfo {author} {\bibfnamefont {Perrin}\ \bibnamefont {Segura}}, \bibinfo
  {author} {\bibfnamefont {Fabian}\ \bibnamefont {Grusdt}}, \bibinfo {author}
  {\bibfnamefont {C{\'e}cile}\ \bibnamefont {Repellin}}, \bibinfo {author}
  {\bibfnamefont {Nathan}\ \bibnamefont {Goldman}}, \ and\ \bibinfo {author}
  {\bibfnamefont {Markus}\ \bibnamefont {Greiner}},\ }\bibfield  {title}
  {\enquote {\bibinfo {title} {Realization of a fractional quantum hall state
  with ultracold atoms},}\ }\href {\doibase 10.1038/s41586-023-06122-4}
  {\bibfield  {journal} {\bibinfo  {journal} {Nature}\ }\textbf {\bibinfo
  {volume} {619}},\ \bibinfo {pages} {495--499} (\bibinfo {year}
  {2023})}\BibitemShut {NoStop}%
\bibitem [{\citenamefont {Bai}\ \emph {et~al.}(2019)\citenamefont {Bai},
  \citenamefont {Bandyopadhyay}, \citenamefont {Pal}, \citenamefont {Suthar},\
  and\ \citenamefont {Angom}}]{bai_19}%
  \BibitemOpen
  \bibfield  {author} {\bibinfo {author} {\bibfnamefont {Rukmani}\ \bibnamefont
  {Bai}}, \bibinfo {author} {\bibfnamefont {Soumik}\ \bibnamefont
  {Bandyopadhyay}}, \bibinfo {author} {\bibfnamefont {Sukla}\ \bibnamefont
  {Pal}}, \bibinfo {author} {\bibfnamefont {K.}~\bibnamefont {Suthar}}, \ and\
  \bibinfo {author} {\bibfnamefont {D.}~\bibnamefont {Angom}},\ }\bibfield
  {title} {\enquote {\bibinfo {title} {Quantum hall states for $\alpha = 1/3$
  in optical lattices},}\ }in\ \href {\doibase 10.1007/978-981-13-9969-5_20}
  {\emph {\bibinfo {booktitle} {Quantum Collisions and Confinement of Atomic
  and Molecular Species, and Photons}}}\ (\bibinfo  {publisher} {Springer
  Singapore},\ \bibinfo {year} {2019})\ pp.\ \bibinfo {pages}
  {211--221}\BibitemShut {NoStop}%
\bibitem [{\citenamefont {Gaur}\ \emph {et~al.}(2023)\citenamefont {Gaur},
  \citenamefont {Sable},\ and\ \citenamefont {Angom}}]{gaur_22}%
  \BibitemOpen
  \bibfield  {author} {\bibinfo {author} {\bibfnamefont {Deepak}\ \bibnamefont
  {Gaur}}, \bibinfo {author} {\bibfnamefont {Hrushikesh}\ \bibnamefont
  {Sable}}, \ and\ \bibinfo {author} {\bibfnamefont {D.}~\bibnamefont
  {Angom}},\ }\bibfield  {title} {\enquote {\bibinfo {title} {Fractional
  quantum hall effect in optical lattices},}\ }\href {\doibase
  10.3389/fphy.2022.1106491} {\bibfield  {journal} {\bibinfo  {journal}
  {Frontiers in Physics}\ }\textbf {\bibinfo {volume} {10}} (\bibinfo {year}
  {2023}),\ 10.3389/fphy.2022.1106491}\BibitemShut {NoStop}%
\bibitem [{\citenamefont {Pal}\ \emph {et~al.}(2019)\citenamefont {Pal},
  \citenamefont {Bai}, \citenamefont {Bandyopadhyay}, \citenamefont {Suthar},\
  and\ \citenamefont {Angom}}]{pal_19}%
  \BibitemOpen
  \bibfield  {author} {\bibinfo {author} {\bibfnamefont {S.}~\bibnamefont
  {Pal}}, \bibinfo {author} {\bibfnamefont {R.}~\bibnamefont {Bai}}, \bibinfo
  {author} {\bibfnamefont {S.}~\bibnamefont {Bandyopadhyay}}, \bibinfo {author}
  {\bibfnamefont {K.}~\bibnamefont {Suthar}}, \ and\ \bibinfo {author}
  {\bibfnamefont {D.}~\bibnamefont {Angom}},\ }\bibfield  {title} {\enquote
  {\bibinfo {title} {Enhancement of the bose glass phase in the presence of an
  artificial gauge field},}\ }\href {\doibase 10.1103/PhysRevA.99.053610}
  {\bibfield  {journal} {\bibinfo  {journal} {Phys. Rev. A}\ }\textbf {\bibinfo
  {volume} {99}},\ \bibinfo {pages} {053610} (\bibinfo {year}
  {2019})}\BibitemShut {NoStop}%
\bibitem [{\citenamefont {Suthar}\ \emph
  {et~al.}(2020{\natexlab{a}})\citenamefont {Suthar}, \citenamefont {Sable},
  \citenamefont {Bai}, \citenamefont {Bandyopadhyay}, \citenamefont {Pal},\
  and\ \citenamefont {Angom}}]{suthar_20}%
  \BibitemOpen
  \bibfield  {author} {\bibinfo {author} {\bibfnamefont {K.}~\bibnamefont
  {Suthar}}, \bibinfo {author} {\bibfnamefont {Hrushikesh}\ \bibnamefont
  {Sable}}, \bibinfo {author} {\bibfnamefont {Rukmani}\ \bibnamefont {Bai}},
  \bibinfo {author} {\bibfnamefont {Soumik}\ \bibnamefont {Bandyopadhyay}},
  \bibinfo {author} {\bibfnamefont {Sukla}\ \bibnamefont {Pal}}, \ and\
  \bibinfo {author} {\bibfnamefont {D.}~\bibnamefont {Angom}},\ }\bibfield
  {title} {\enquote {\bibinfo {title} {Supersolid phase of the extended
  bose-hubbard model with an artificial gauge field},}\ }\href {\doibase
  10.1103/PhysRevA.102.013320} {\bibfield  {journal} {\bibinfo  {journal}
  {Phys. Rev. A}\ }\textbf {\bibinfo {volume} {102}},\ \bibinfo {pages}
  {013320} (\bibinfo {year} {2020}{\natexlab{a}})}\BibitemShut {NoStop}%
\bibitem [{\citenamefont {Suthar}\ \emph
  {et~al.}(2020{\natexlab{b}})\citenamefont {Suthar}, \citenamefont {Kraus},
  \citenamefont {Sable}, \citenamefont {Angom}, \citenamefont {Morigi},\ and\
  \citenamefont {Zakrzewski}}]{suthar_20_2}%
  \BibitemOpen
  \bibfield  {author} {\bibinfo {author} {\bibfnamefont {Kuldeep}\ \bibnamefont
  {Suthar}}, \bibinfo {author} {\bibfnamefont {Rebecca}\ \bibnamefont {Kraus}},
  \bibinfo {author} {\bibfnamefont {Hrushikesh}\ \bibnamefont {Sable}},
  \bibinfo {author} {\bibfnamefont {Dilip}\ \bibnamefont {Angom}}, \bibinfo
  {author} {\bibfnamefont {Giovanna}\ \bibnamefont {Morigi}}, \ and\ \bibinfo
  {author} {\bibfnamefont {Jakub}\ \bibnamefont {Zakrzewski}},\ }\bibfield
  {title} {\enquote {\bibinfo {title} {Staggered superfluid phases of dipolar
  bosons in two-dimensional square lattices},}\ }\href {\doibase
  10.1103/PhysRevB.102.214503} {\bibfield  {journal} {\bibinfo  {journal}
  {Phys. Rev. B}\ }\textbf {\bibinfo {volume} {102}},\ \bibinfo {pages}
  {214503} (\bibinfo {year} {2020}{\natexlab{b}})}\BibitemShut {NoStop}%
\bibitem [{\citenamefont {Bai}\ \emph {et~al.}(2020)\citenamefont {Bai},
  \citenamefont {Gaur}, \citenamefont {Sable}, \citenamefont {Bandyopadhyay},
  \citenamefont {Suthar},\ and\ \citenamefont {Angom}}]{bai_20}%
  \BibitemOpen
  \bibfield  {author} {\bibinfo {author} {\bibfnamefont {Rukmani}\ \bibnamefont
  {Bai}}, \bibinfo {author} {\bibfnamefont {Deepak}\ \bibnamefont {Gaur}},
  \bibinfo {author} {\bibfnamefont {Hrushikesh}\ \bibnamefont {Sable}},
  \bibinfo {author} {\bibfnamefont {Soumik}\ \bibnamefont {Bandyopadhyay}},
  \bibinfo {author} {\bibfnamefont {K.}~\bibnamefont {Suthar}}, \ and\ \bibinfo
  {author} {\bibfnamefont {D.}~\bibnamefont {Angom}},\ }\bibfield  {title}
  {\enquote {\bibinfo {title} {Segregated quantum phases of dipolar bosonic
  mixtures in two-dimensional optical lattices},}\ }\href {\doibase
  10.1103/PhysRevA.102.043309} {\bibfield  {journal} {\bibinfo  {journal}
  {Phys. Rev. A}\ }\textbf {\bibinfo {volume} {102}},\ \bibinfo {pages}
  {043309} (\bibinfo {year} {2020})}\BibitemShut {NoStop}%
\bibitem [{\citenamefont {Bandyopadhyay}\ \emph {et~al.}(2022)\citenamefont
  {Bandyopadhyay}, \citenamefont {Sable}, \citenamefont {Gaur}, \citenamefont
  {Bai}, \citenamefont {Mukerjee},\ and\ \citenamefont
  {Angom}}]{bandyopadhyay_22}%
  \BibitemOpen
  \bibfield  {author} {\bibinfo {author} {\bibfnamefont {Soumik}\ \bibnamefont
  {Bandyopadhyay}}, \bibinfo {author} {\bibfnamefont {Hrushikesh}\ \bibnamefont
  {Sable}}, \bibinfo {author} {\bibfnamefont {Deepak}\ \bibnamefont {Gaur}},
  \bibinfo {author} {\bibfnamefont {Rukmani}\ \bibnamefont {Bai}}, \bibinfo
  {author} {\bibfnamefont {Subroto}\ \bibnamefont {Mukerjee}}, \ and\ \bibinfo
  {author} {\bibfnamefont {D.}~\bibnamefont {Angom}},\ }\bibfield  {title}
  {\enquote {\bibinfo {title} {Quantum phases of dipolar bosons in a multilayer
  optical lattice},}\ }\href {\doibase 10.1103/PhysRevA.106.043301} {\bibfield
  {journal} {\bibinfo  {journal} {Phys. Rev. A}\ }\textbf {\bibinfo {volume}
  {106}},\ \bibinfo {pages} {043301} (\bibinfo {year} {2022})}\BibitemShut
  {NoStop}%
\bibitem [{\citenamefont {Kitaev}\ and\ \citenamefont
  {Preskill}(2006)}]{kitaev_06}%
  \BibitemOpen
  \bibfield  {author} {\bibinfo {author} {\bibfnamefont {Alexei}\ \bibnamefont
  {Kitaev}}\ and\ \bibinfo {author} {\bibfnamefont {John}\ \bibnamefont
  {Preskill}},\ }\bibfield  {title} {\enquote {\bibinfo {title} {Topological
  entanglement entropy},}\ }\href {\doibase 10.1103/PhysRevLett.96.110404}
  {\bibfield  {journal} {\bibinfo  {journal} {Phys. Rev. Lett.}\ }\textbf
  {\bibinfo {volume} {96}},\ \bibinfo {pages} {110404} (\bibinfo {year}
  {2006})}\BibitemShut {NoStop}%
\bibitem [{\citenamefont {Levin}\ and\ \citenamefont {Wen}(2006)}]{levin_06}%
  \BibitemOpen
  \bibfield  {author} {\bibinfo {author} {\bibfnamefont {Michael}\ \bibnamefont
  {Levin}}\ and\ \bibinfo {author} {\bibfnamefont {Xiao-Gang}\ \bibnamefont
  {Wen}},\ }\bibfield  {title} {\enquote {\bibinfo {title} {Detecting
  topological order in a ground state wave function},}\ }\href {\doibase
  10.1103/PhysRevLett.96.110405} {\bibfield  {journal} {\bibinfo  {journal}
  {Phys. Rev. Lett.}\ }\textbf {\bibinfo {volume} {96}},\ \bibinfo {pages}
  {110405} (\bibinfo {year} {2006})}\BibitemShut {NoStop}%
\bibitem [{\citenamefont {Krutitsky}(2016)}]{krutitsky_16}%
  \BibitemOpen
  \bibfield  {author} {\bibinfo {author} {\bibfnamefont {Konstantin~V.}\
  \bibnamefont {Krutitsky}},\ }\bibfield  {title} {\enquote {\bibinfo {title}
  {Ultracold bosons with short-range interaction in regular optical
  lattices},}\ }\href {\doibase https://doi.org/10.1016/j.physrep.2015.10.004}
  {\bibfield  {journal} {\bibinfo  {journal} {Physics Reports}\ }\textbf
  {\bibinfo {volume} {607}},\ \bibinfo {pages} {1--101} (\bibinfo {year}
  {2016})},\ \bibinfo {note} {ultracold bosons with short-range interaction in
  regular optical lattices}\BibitemShut {NoStop}%
\bibitem [{\citenamefont {L\"uhmann}(2013)}]{luhmann_13}%
  \BibitemOpen
  \bibfield  {author} {\bibinfo {author} {\bibfnamefont {D.-S.}\ \bibnamefont
  {L\"uhmann}},\ }\bibfield  {title} {\enquote {\bibinfo {title} {Cluster
  {G}utzwiller method for bosonic lattice systems},}\ }\href {\doibase
  10.1103/PhysRevA.87.043619} {\bibfield  {journal} {\bibinfo  {journal} {Phys.
  Rev. A}\ }\textbf {\bibinfo {volume} {87}},\ \bibinfo {pages} {043619}
  (\bibinfo {year} {2013})}\BibitemShut {NoStop}%
\bibitem [{\citenamefont {Bai}(2018)}]{bai_thesis}%
  \BibitemOpen
  \bibfield  {author} {\bibinfo {author} {\bibfnamefont {Rukmani}\ \bibnamefont
  {Bai}},\ }\emph {\bibinfo {title} {Synthetic Magnetic Fields and
  Multi-Component BECs in Optical Lattices}},\ \href@noop {} {Ph.D. thesis},\
  \bibinfo  {school} {Indian Institute of Technology Gandhinagar} (\bibinfo
  {year} {2018})\BibitemShut {NoStop}%
\bibitem [{\citenamefont {Capogrosso-Sansone}\ \emph
  {et~al.}(2008)\citenamefont {Capogrosso-Sansone}, \citenamefont {S\"oyler},
  \citenamefont {Prokof'ev},\ and\ \citenamefont {Svistunov}}]{cappogrosso_08}%
  \BibitemOpen
  \bibfield  {author} {\bibinfo {author} {\bibfnamefont {Barbara}\ \bibnamefont
  {Capogrosso-Sansone}}, \bibinfo {author} {\bibfnamefont {\ifmmode
  \mbox{\c{S}}\else \c{S}\fi{}ebnem G\"une\ifmmode
  \mbox{\c{s}}\else~\c{s}\fi{}}\ \bibnamefont {S\"oyler}}, \bibinfo {author}
  {\bibfnamefont {Nikolay}\ \bibnamefont {Prokof'ev}}, \ and\ \bibinfo {author}
  {\bibfnamefont {Boris}\ \bibnamefont {Svistunov}},\ }\bibfield  {title}
  {\enquote {\bibinfo {title} {Monte carlo study of the two-dimensional
  bose-hubbard model},}\ }\href {\doibase 10.1103/PhysRevA.77.015602}
  {\bibfield  {journal} {\bibinfo  {journal} {Phys. Rev. A}\ }\textbf {\bibinfo
  {volume} {77}} (\bibinfo {year} {2008}),\
  10.1103/PhysRevA.77.015602}\BibitemShut {NoStop}%
\bibitem [{qmc()}]{qmc_comm}%
  \BibitemOpen
  \href@noop {} {}\bibinfo {note} {{ The QMC data is obtained from personal
  communication with Prof. Barbara Capogrosso-Sansone.}}\BibitemShut {Stop}%
\bibitem [{\citenamefont {Wang}()}]{zhentao_wang}%
  \BibitemOpen
  \bibfield  {author} {\bibinfo {author} {\bibfnamefont {Zhentao}\ \bibnamefont
  {Wang}},\ }\href {https://github.com/wztzjhn/quantum_basis} {\enquote
  {\bibinfo {title} {Quantum basis: Exact diagonalization library for general
  models, https://github.com/wztzjhn/quantum\_basis},}\ }\BibitemShut {NoStop}%
\bibitem [{\citenamefont {Ekert}\ and\ \citenamefont
  {Knight}(1995)}]{ekert_95}%
  \BibitemOpen
  \bibfield  {author} {\bibinfo {author} {\bibfnamefont {Artur}\ \bibnamefont
  {Ekert}}\ and\ \bibinfo {author} {\bibfnamefont {Peter~L.}\ \bibnamefont
  {Knight}},\ }\bibfield  {title} {\enquote {\bibinfo {title} {{Entangled
  quantum systems and the Schmidt decomposition}},}\ }\href {\doibase
  10.1119/1.17904} {\bibfield  {journal} {\bibinfo  {journal} {American Journal
  of Physics}\ }\textbf {\bibinfo {volume} {63}},\ \bibinfo {pages} {415--423}
  (\bibinfo {year} {1995})},\ \Eprint
  {http://arxiv.org/abs/https://pubs.aip.org/aapt/ajp/article-pdf/63/5/415/11806926/415\_1\_online.pdf}
  {https://pubs.aip.org/aapt/ajp/article-pdf/63/5/415/11806926/415\_1\_online.pdf}
  \BibitemShut {NoStop}%
\bibitem [{\citenamefont {Eisert}\ \emph {et~al.}(2010)\citenamefont {Eisert},
  \citenamefont {Cramer},\ and\ \citenamefont {Plenio}}]{eisert_10}%
  \BibitemOpen
  \bibfield  {author} {\bibinfo {author} {\bibfnamefont {J.}~\bibnamefont
  {Eisert}}, \bibinfo {author} {\bibfnamefont {M.}~\bibnamefont {Cramer}}, \
  and\ \bibinfo {author} {\bibfnamefont {M.~B.}\ \bibnamefont {Plenio}},\
  }\bibfield  {title} {\enquote {\bibinfo {title} {Colloquium: Area laws for
  the entanglement entropy},}\ }\href {\doibase 10.1103/RevModPhys.82.277}
  {\bibfield  {journal} {\bibinfo  {journal} {Rev. Mod. Phys.}\ }\textbf
  {\bibinfo {volume} {82}},\ \bibinfo {pages} {277--306} (\bibinfo {year}
  {2010})}\BibitemShut {NoStop}%
\bibitem [{\citenamefont {Zhang}\ and\ \citenamefont {Liu}(2017)}]{zhang_17}%
  \BibitemOpen
  \bibfield  {author} {\bibinfo {author} {\bibfnamefont {Jiang-Min}\
  \bibnamefont {Zhang}}\ and\ \bibinfo {author} {\bibfnamefont
  {Yu}~\bibnamefont {Liu}},\ }\bibfield  {title} {\enquote {\bibinfo {title}
  {Geometric entanglement in the laughlin wave function},}\ }\href {\doibase
  10.1088/1367-2630/aa7e72} {\bibfield  {journal} {\bibinfo  {journal} {New
  Journal of Physics}\ }\textbf {\bibinfo {volume} {19}},\ \bibinfo {pages}
  {083019} (\bibinfo {year} {2017})}\BibitemShut {NoStop}%
\bibitem [{\citenamefont {Jaksch}\ and\ \citenamefont
  {Zoller}(2003)}]{jaksch_03}%
  \BibitemOpen
  \bibfield  {author} {\bibinfo {author} {\bibfnamefont {D.}~\bibnamefont
  {Jaksch}}\ and\ \bibinfo {author} {\bibfnamefont {P.}~\bibnamefont
  {Zoller}},\ }\bibfield  {title} {\enquote {\bibinfo {title} {Creation of
  effective magnetic fields in optical lattices: the {H}ofstadter butterfly for
  cold neutral atoms},}\ }\href {\doibase
  https://doi.org/10.1088/1367-2630/5/1/356} {\bibfield  {journal} {\bibinfo
  {journal} {New J. Phys.}\ }\textbf {\bibinfo {volume} {5}},\ \bibinfo {pages}
  {56} (\bibinfo {year} {2003})}\BibitemShut {NoStop}%
\bibitem [{\citenamefont {Wen}\ and\ \citenamefont {Niu}(1990)}]{wen_90}%
  \BibitemOpen
  \bibfield  {author} {\bibinfo {author} {\bibfnamefont {X.~G.}\ \bibnamefont
  {Wen}}\ and\ \bibinfo {author} {\bibfnamefont {Q.}~\bibnamefont {Niu}},\
  }\bibfield  {title} {\enquote {\bibinfo {title} {Ground-state degeneracy of
  the fractional quantum hall states in the presence of a random potential and
  on high-genus riemann surfaces},}\ }\href {\doibase 10.1103/PhysRevB.41.9377}
  {\bibfield  {journal} {\bibinfo  {journal} {Phys. Rev. B}\ }\textbf {\bibinfo
  {volume} {41}},\ \bibinfo {pages} {9377--9396} (\bibinfo {year}
  {1990})}\BibitemShut {NoStop}%
\bibitem [{\citenamefont {Hatsugai}(2005)}]{hatsugai_05}%
  \BibitemOpen
  \bibfield  {author} {\bibinfo {author} {\bibfnamefont {Yasuhiro}\
  \bibnamefont {Hatsugai}},\ }\bibfield  {title} {\enquote {\bibinfo {title}
  {Characterization of topological insulators: Chern numbers for ground state
  multiplet},}\ }\href {\doibase 10.1143/JPSJ.74.1374} {\bibfield  {journal}
  {\bibinfo  {journal} {Journal of the Physical Society of Japan}\ }\textbf
  {\bibinfo {volume} {74}},\ \bibinfo {pages} {1374--1377} (\bibinfo {year}
  {2005})},\ \Eprint
  {http://arxiv.org/abs/https://doi.org/10.1143/JPSJ.74.1374}
  {https://doi.org/10.1143/JPSJ.74.1374} \BibitemShut {NoStop}%
\bibitem [{\citenamefont {Maschhoff}\ and\ \citenamefont
  {Sorensen}(1996)}]{maschhoff_96}%
  \BibitemOpen
  \bibfield  {author} {\bibinfo {author} {\bibfnamefont {K.~J.}\ \bibnamefont
  {Maschhoff}}\ and\ \bibinfo {author} {\bibfnamefont {D.~C.}\ \bibnamefont
  {Sorensen}},\ }\bibfield  {title} {\enquote {\bibinfo {title} {P{\_}arpack:
  An efficient portable large scale eigenvalue package for distributed memory
  parallel architectures},}\ }in\ \href@noop {} {\emph {\bibinfo {booktitle}
  {Applied Parallel Computing Industrial Computation and Optimization}}},\
  \bibinfo {editor} {edited by\ \bibinfo {editor} {\bibfnamefont {Jerzy}\
  \bibnamefont {Wa{\'{s}}niewski}}, \bibinfo {editor} {\bibfnamefont {Jack}\
  \bibnamefont {Dongarra}}, \bibinfo {editor} {\bibfnamefont {Kaj}\
  \bibnamefont {Madsen}}, \ and\ \bibinfo {editor} {\bibfnamefont {Dorte}\
  \bibnamefont {Olesen}}}\ (\bibinfo  {publisher} {Springer Berlin
  Heidelberg},\ \bibinfo {address} {Berlin, Heidelberg},\ \bibinfo {year}
  {1996})\ pp.\ \bibinfo {pages} {478--486}\BibitemShut {NoStop}%
\end{thebibliography}%

\end{document}